# A Scalable Synthesis Algorithm for Reversible Functions

Moein Sarvaghad-Moghaddam, Morteza Saheb Zamani[*], Mehdi Sedighi

Quantum Design Automation Laboratory, Department of Computer Engineering, Amirkabir University of Technology, Tehran, Iran.

**Abstract**: Reversible computation is an emerging technology that has gained significant attention due to its critical role in quantum circuit synthesis and low-power design. This paper introduces a transformation-based method for exact synthesis of reversible circuits. The proposed approach utilizes a novel adaptation of the Quine-McCluskey algorithm to eliminate input-output discrepancies in the truth table, transforming the permutation matrix into an identity matrix. Furthermore, a novel search space reduction technique is presented which, combined with the primary method, enables the synthesis algorithm to handle high-input reversible functions. This approach combines the influence of multiple control qubits on a target qubit, evaluating their collective impact. This aggregation can decrease the control qubit count within quantum gates. Consequently, it proves beneficial for applications like surface code error correction architectures as well as current Noisy Intermediate-Scale Quantum (NISQ) hardwares. Experimental results demonstrate significant improvements over the state-of-the-art exact synthesis methods, achieving up to 99% improvements in terms of the number of levels of T-gates.

## 1. Introduction

Quantum computing [1] has been gaining traction due to its potential to solve complex problems using algorithms such as Shor's algorithm [2] for finding the prime factors of an integer and Grover's algorithm [3] as a quantum search method.

Significant progress has been made by major companies like IBM and Google in the field of quantum computing. However, there are still limitations such as qubit noise, short coherence time, and the inability to fully control qubits, which make it challenging to build a practical quantum computer. These constraints limit the number of qubits that can be integrated into a single chip to a certain threshold. Moreover, to reduce errors, it is essential to include a quantum error correction module, such as surface codes. However, integrating this module into a quantum computer significantly increases the number of additional qubits needed. To tackle these challenges, it is crucial to design a circuit that uses a smaller number of quantum gates while ensuring that no ancilla qubits are required.

---
[*] Corresponding author
Email: szamani@aut.ac.ir

To design a quantum circuit that can be implemented on a quantum platform, synthesis algorithms are required. The process of quantum synthesis begins with an initial matrix known as the *transformation matrix*, which describes the functionality of the circuit. Next, the transformation matrix is converted into a quantum circuit using a library of quantum gates. Then, the quantum circuit generated from the previous step is mapped onto a circuit containing only the components available in the target library of gates.

Reversible circuit synthesis is used as an important stage in the process of some quantum circuit synthesis algorithms [4,5]. This stage occurs after removing the superposition from the transformation matrix. The resulting matrix is known as a *permutation matrix* with entries consisting solely of 0s and 1s. The transform function of the circuit can also be specified by a truth table like the one in logic design with $n$ inputs ($x_1, x_2, ..., x_n$) and $n$ outputs ($f_1, f_2, ..., f_n$) such that the mapping of inputs to the outputs is unique. Furthermore, reversible circuits have limitations, such as the lack of feedback and fanout, leading to the need for synthesis algorithms specifically designed for them.

The NCT library, which includes NOT, CNOT, and Toffoli gates, is usually used for reversible circuit synthesis. Multi-Controlled Toffoli (MCT) gates may also be required to facilitate the synthesis process.

In quantum computing, due to the mentioned limitations, circuits with fewer control inputs and intermediary gates lead to a reduction in the depth of the circuit as well as a reduction in output errors. On the other hand, the Clifford+T library is one of the well-known fault-tolerant quantum libraries used on existing quantum hardware. Considering that the expensive T gates are required to implement multi-qubit controlled gates (especially MCT gates), reducing the number of control qubits of the controlled gates leads to a further reduction of T gates in the final circuit.

Reversible synthesis methods can be divided into two categories. The first category includes exact methods, such as search and lookup algorithms [6] and satisfiability (SAT)-based methods [7, 8], which aim for the absolute optimal solution, but due to the complexity of the problem, it can only operate on a small number of qubits. The second category includes heuristic methods such as [9, 10]. Even though they can be applied to more qubits, they do not guarantee the optimal solution and their cost can be improved.

Various methods have been proposed to improve the synthesis of reversible circuits, including cycle-based [11-13], block-based [14], transformation-based [15], exclusive sum of product (ESOP) based [16–20], decision diagram-based methods such as binary decision diagrams (BDD) [21], functional decision diagrams (FDD) [22], Kronecker FDD (KFDD) [23], and quantum multiple-valued decision diagrams (QMDD) [24]. Moreover, exploiting the data structures BDDs/QMDDs [25] and also structural synthesis methods of ESOP [19, 26] are known to be effective techniques for synthesizing large functions. However, the generated circuits require a large number of ancilla lines and there is still room for further improvement in the results.

The authors in [4] introduced a synthesis method for quantum circuits based on transformation. They utilize a global view approach, which involves assessing the impact of all superpositions related to Hadamard gates in the transformation matrix to minimize the control qubits of this gate. This approach is specifically applied to the H gate, and only works for certain transformation matrices. When the conditions are not met, their method reverts to a local view method presented in [5], which involves checking elements column-wise in the permutation matrix.

In this paper, a transformation-based synthesis method for reversible circuits is presented. The library considered in this paper is the NCT library. We presented a novel adaptation of the Quine McCluskey (QM) method, which is a classical and traditional method for logic circuit optimization, specifically designed for synthesizing reversible circuits. The proposed approach considers all control qubits associated with control gates that impact each input, instead of considering the gates locally. This allows us to design gates with much fewer control qubits. Moreover, this method does not require ancilla qubits, and the outputs are generated directly from the input qubits. This provides a significant advantage for current quantum hardware in the NISQ era with quantum error correction modules. Then, a search space compression method is introduced to synthesize circuits with a large number of inputs, allowing the proposed method to be applied to circuits with more inputs within a substantially smaller search space.

The rest of the paper is organized as follows. Section 2 presents some related background about quantum gates and the QM method. In Section 3, the general idea of our method is explained. Section 4 describes the proposed method in detail. Section 5 introduces a method that enhances the scalability of our approach, enabling it to efficiently handle functions with a larger number of variables. Section 6 presents the experimental results. Finally, Section 7 concludes the paper.

2. **Basic concepts**

- **Quantum gates**

One of the well-known libraries for synthesizing reversible circuits is the NCT library. In this paper, we consider it as the basic library. It includes $C^m NOT$ gates. A $C^m NOT(x_1, x_2, \ldots, x_m, x_{m+1})$ gate is a gate with $m$ control qubits and one qubit, $x_{m+1}$, as the target qubit. In particular, $C^0 NOT, C^1 NOT, C^2 NOT$ gates are referred to as $NOT, CNOT$ and *Toffoli* gates, respectively. In this paper, we represent Toffoli gates with $n$ qubits ($n \geq$

3) by $Toffli\ n$. For example, $Toffli\ 4$ means a Toffoli gate with four control qubits $(x_1, x_2, x_3, x_4)$ and one target qubit $x_5$ as $Toff(x_1, x_2, x_3, x_4, x_5)$.

- **Quine-McCluskey method**

In the traditional Boolean logic design, Karnaugh maps (K-maps) and Quine-McCluskey (QM) Methods are two well-known methods for generating optimized circuits. The K-map method is suitable for manual optimization of functions with up to 5 inputs, but it provides an excellent graphical representation that makes the simplification process easy to comprehend. The second method is systematic and tabulation-based, allowing it to work for more than 5 inputs and complex functions and making it suitable for computer-based simplification.

In the rest of the paper, we present our method based on the QM method but utilize K-maps for graphical representation to clearly explain the concepts of our algorithm.

Both K-map and QM tabular methods use three concepts to simplify Boolean functions. These concepts are (a) *implicant* which means a group of $2^i$ adjacent minterms with the value of 1 where $i$ can be an integer number in $[0, n]$ for an $n$-variable function; (b) *prime implicant* (*PI*) which is the largest possible implicants; and (3) *essential prime implicants* (*EPI*) which are the PIs that cover at least one minterm which cannot be covered by other PIs.

The steps of the QM method can be summarized as follows:

1. Arrange the given minterms into groups in ascending order based on the number of ones in their binary representation.
2. Compare the minterms placed in one group with the minterms placed in consecutive groups in pairs. If there is a Hamming distance of one between two minterms (*i.e.*, a difference in one bit), then combine the two minterms by replacing the differing bit with an X, and keeping the rest of the bits the same. Mark any two minterms that can be combined as visited.
3. Repeat the second step for new terms until no implicant combines with another. The unmarked terms are all PIs of the function.
4. Create a PI table with the PIs in the rows and the minterms in the columns. Then, assign a value of 1 to each table's cell that corresponds to a minterm covered by a PI.
5. The columns of the table above are checked. If a minterm is covered by only one PI, then that PI is considered an EPI, which is added as a term in the simplified Boolean expression.
6. Remove rows and columns related to all EPIs.
7. Cover the rest of minterms with a minimum number of PIs.

## 3. General idea of the proposed approach

Transformation-based methods for synthesizing reversible circuits from a truth table attempt to find a way to convert the inputs to the outputs of the function represented by the table. This goal can be achieved by eliminating the differences between the input and output variables in the truth table through a series of reversible operations (gates). Then, by applying the conjugate transpose of these operations, we can effectively convert the inputs to the desired outputs.

In the proposed method, instead of working directly with the inputs ($x_i$) and the outputs ($f_i$), for each variable $x_i$, we obtain a function $v_i = x_i \oplus f_i$ and then try to change the 1 values of this function to 0 by applying a sequence of reversible gates. This will make the input $x_i$ equal to $f_i$ step by step. For example, Figure 1 shows the truth table for a four-variable reversible function. The inputs ($x_i$) with values different from the corresponding outputs ($f_i$) are shown in circles. The last four columns show the truth table for the $v_i$ functions in which 1's in each column corresponds to a circle in the input column ($x_i$).

| $x_1$ | $x_2$ | $x_3$ | $x_4$ | $f_1$ | $f_2$ | $f_3$ | $f_4$ | $v_1$ | $v_2$ | $v_3$ | $v_4$ |
|---|---|---|---|---|---|---|---|---|---|---|---|
| 0 | 0 | 0 | (0) | 0 | 0 | 0 | 1 | 0 | 0 | 0 | 1 |
| 0 | (0) | 0 | 1 | 0 | 1 | 0 | 1 | 0 | 1 | 0 | 0 |
| 0 | 0 | (1) | 0 | 0 | 0 | 0 | 0 | 0 | 0 | 1 | 0 |
| (0) | 0 | (1) | (1) | 1 | 0 | 0 | 0 | 1 | 0 | 1 | 1 |
| (0) | (1) | 0 | (0) | 1 | 0 | 0 | 1 | 1 | 1 | 0 | 1 |
| (0) | (1) | (0) | 1 | 1 | 0 | 1 | 1 | 1 | 1 | 1 | 0 |
| 0 | (1) | 1 | 0 | 0 | 0 | 1 | 0 | 0 | 1 | 0 | 0 |
| (0) | 1 | 1 | 1 | 1 | 1 | 1 | 1 | 1 | 0 | 0 | 0 |
| (1) | 0 | (0) | (0) | 0 | 0 | 1 | 1 | 1 | 0 | 1 | 1 |
| 1 | (0) | 0 | (1) | 1 | 1 | 0 | 0 | 0 | 1 | 0 | 1 |
| (1) | (0) | (1) | 0 | 0 | 1 | 0 | 0 | 1 | 1 | 1 | 0 |
| (1) | (0) | 1 | (1) | 0 | 1 | 1 | 0 | 1 | 1 | 0 | 1 |
| 1 | (1) | (0) | 0 | 1 | 0 | 1 | 0 | 0 | 1 | 1 | 0 |
| 1 | 1 | (0) | (1) | 1 | 1 | 1 | 0 | 0 | 0 | 1 | 1 |
| 1 | 1 | (1) | (0) | 1 | 1 | 0 | 1 | 0 | 0 | 1 | 1 |
| (1) | 1 | 1 | 1 | 0 | 1 | 1 | 1 | 1 | 0 | 0 | 0 |

Figure 1: Illustration of truth table of reversible function with vector $v_i$.

The proposed synthesis algorithm uses a set of rules based on specific templates on the binary representation of functions, minterms and PIs. However, before presenting the algorithm, the basic methods used in the algorithm are explained here by examples.

To synthesize the circuit with lower gate costs, we can use a K-map for elements with 1 value of each $v_i$ function. For the function $v_i$, each PI in the K-map corresponds to a CNOT gate with the PI's product term as its control qubit and the variable $x_i$ (called *main variable* herein) as its target qubit (i.e., as an ESOP expression $PI \oplus x_i$). By applying this gate, we can eliminate the 1's in the map, thereby making the target output equal to the input. Figure 2.a shows the K-map of the function $v_4$ with a PI of two minterms with a 1 value, which are converted to zero by applying the $CNOT((x_1'.x_2.x_3), x_4)$ gate[†]. On the other hand, larger PIs (*i.e.,* those with more 1's in the K-map) correspond to CNOT gates with fewer control qubits. Therefore, our method aims to cover minterms with larger PIs to remove more 1's of $v_i$ with smaller CNOT gates (see Figure 2.b).

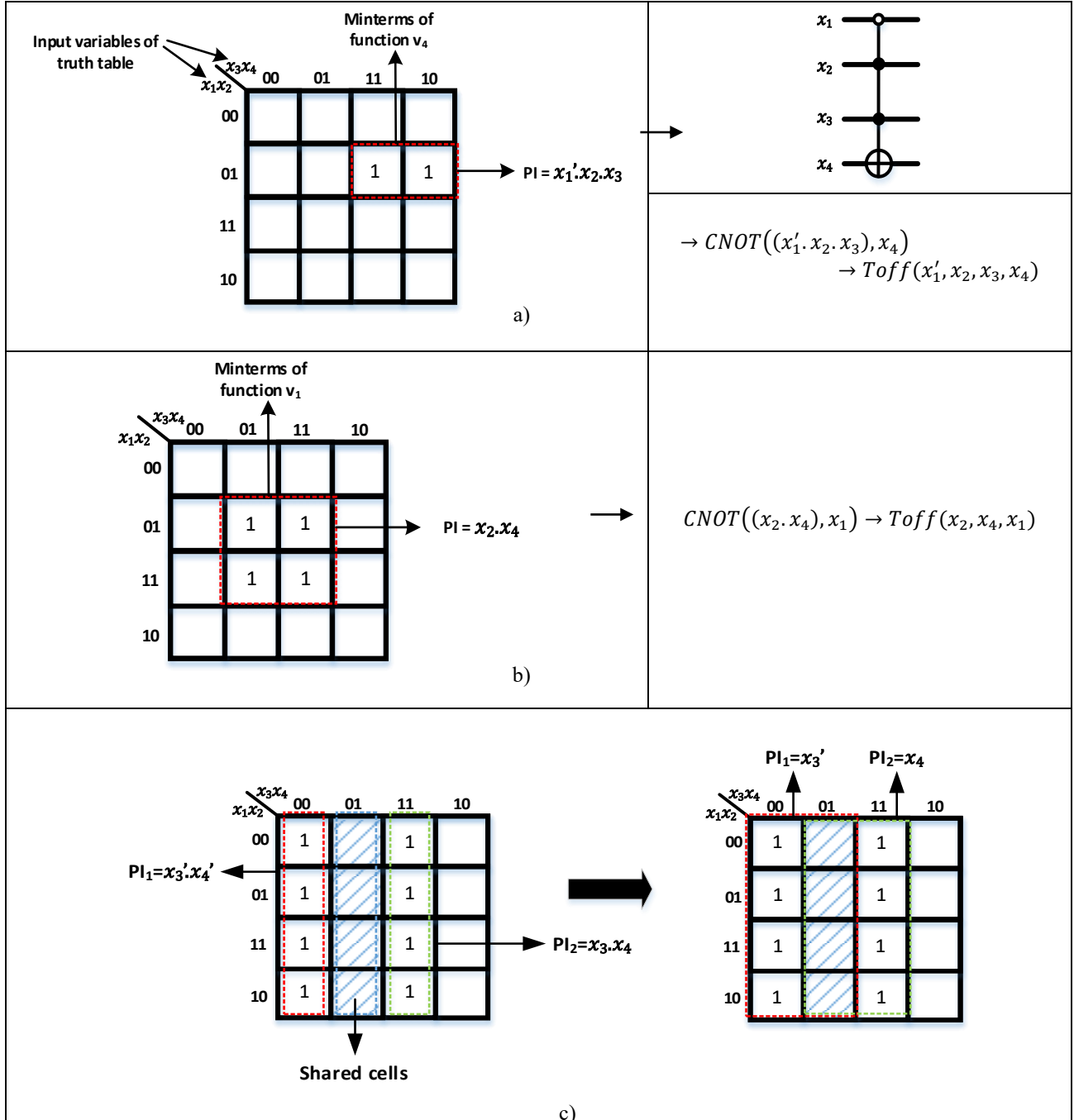

Figure 2: illustration of some rules used in the proposed method

---

[†] The K-maps in Figure 2 do not necessarily correspond to the function in Figure 1.

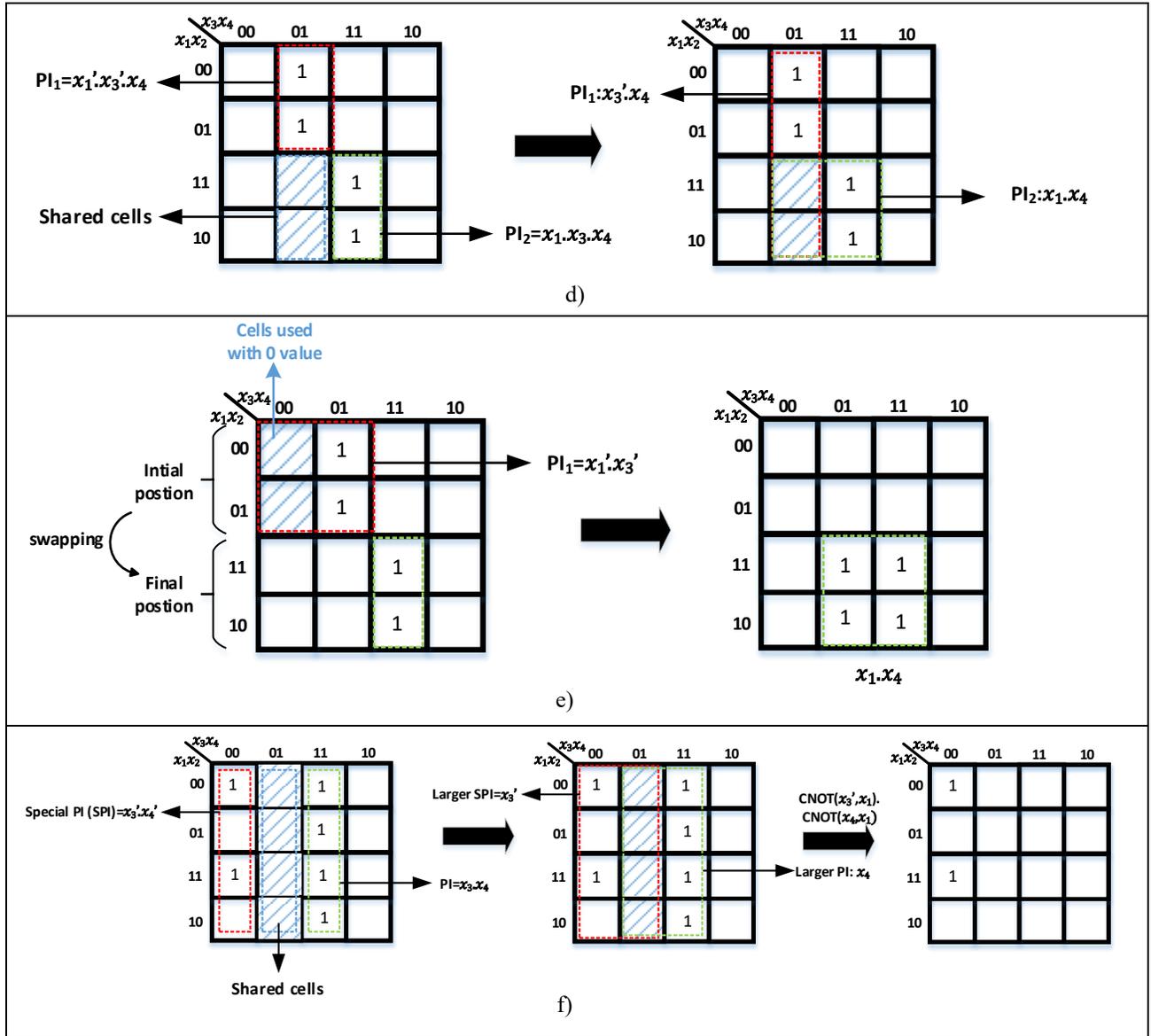

Figure 3: illustration of some rules used in the proposed method (continued)

In the general case, when there are several disjoint PIs in a K-map ($PI_j$, $j \in [1, p]$ where $p$ is the number of disjoint PIs), the circuit will contain a CNOT gate with ESOP expression $\oplus_j (PI_j)$ as its control qubit and the main variable as its target qubit[‡]. Therefore, for the main variable $x_i$, the disjoint PIs lead to the following expression:

$$[\oplus_j(PI_j)] \oplus x_i \rightarrow CNOT(\oplus_j (PI_j), x_i) \qquad (1)$$

---

[‡] We use the XOR symbol for ESOP expressions as sigma is used for SOP expressions.

The part $[\oplus_j(PI_j)]$ in the above expression can be potentially simplified by factoring in Boolean algebra. On the other hand, given that the CNOT function is linear and the target of the CNOT gates is the same, we can rewrite the above expression as follows:

$$[\oplus_j(PI_j \oplus x_i)] \qquad (2)$$

which can be implemented as the following:

$$\prod_j CNOT(PI_j, x_i) \qquad (3)$$

We can use Eq.3 instead of Eq.2 if no common factor exists for factoring among the PIs.

Let's consider the PIs in Figure 2.c for the function $v_1$ with the main variable $x_1$. Then, the product terms $PI_1$ and $PI_2$ are $x_3'.x_4'$ and $x_3.x_4$, respectively, and according to Eq.1, the function can be stated as the following ESOP expression:

$$[x_3'.x_4' \oplus x_3.x_4] \oplus x_1$$

We can simplify the expression in the bracket as follows:

$$[x_3' \oplus x_4] \oplus x_1$$

Then, according to Eq.2 and Eq.3, we can implement the above equation using two parallel CNOT gates as the following:

$$CNOT((x_3'.x_4' \oplus x_3.x_4), x_1) = CNOT((x_3' \oplus x_4), x_1) = CNOT(x_3', x_1).CNOT(x_4, x_1)$$

From another point of view, we will illustrate the above operations by K-map using shared cells. If a cell in the K-map is used twice, shared between two PIs, it does not affect the function $v_i$ (since $1 \oplus 1 = 0$). This feature can be useful when a larger PI is created via shared cells merged with each PI. Let us revisit Figure 2.c, which presents a K-map for $v_1$. It contains two PIs corresponding to the product terms of $x_3'.x_4'$ and $x_3.x_4$. For instance, if cells with the product term $x_3'.x_4$ are considered as the XOR of two 1's, they can be shared by larger PIs, namely $x_3'$ and $x_4$. Finally, $CNOT(x_3', x_1).CNOT(x_4, x_1)$ can be applied to remove 1's from the K-map.

In some cases, swapping the 1's of an implicant with a set of neighboring cells in the K-map can place those 1's next to another implicant, leading to a larger PI. The 1's of an implicant can be swapped with the neighboring cells using a CNOT gate in which the target qubit is not the main variable. In this process, we can use and merge the free neighboring cells with 1's of the implicant to reduce the number of control qubits of the CNOT gate.

For a main variable $x_i$, let's label the implicant with a group of 1's as $P_1$ and the neighboring group as $P_2$. The position of the two groups is swapped by a CNOT gate with the common term of the two mentioned positions as the control qubit and the qubit which is changed for the two positions as the target qubit. Since the target qubit is different from the main variable of $x_i$, this CNOT gate will only result in the swapping of cells of the two groups within the K-map. It is important to note that when the CNOT gate is applied to a target qubit (for example, $x_2$), and that qubit is not the main variable in the K-map, it will affect its corresponding K-map, where it is the main variable. This will lead to a change in the ones in that K-map. In this case, if the function $v_2$ is zero, a CNOT gate should be applied after synthesizing the function of the above variable to cancel out its effect on $v_2$. If the function $v_2$ is not zero, these effects will be considered during synthesizing the function $v_2$.

For example, Figure 2.d illustrates the K-map for variable $x_2$. There are two PIs (PI$_1$ = $x_1'.x_3'.x_4$ and PI$_2$ = $x_1.x_3.x_4$) in this K-map, which generates two consecutive gates $Toff(x_1', x_3', x_4, x_2)$ and $Toff(x_1, x_3, x_4, x_2)$. We can synthesize the function for $x_2$ through two processes. First, we can consider the cells with a product term $x_1.x_3'.x_4$ and share them with the two above PIs to create two larger PIs, namely $x_3'.x_4$ and $x_1.x_4$. In this case, the ESOP term will be $(x_3'.x_4 \oplus x_1.x_4) \oplus x_2$. Then, we need to simplify and factorize the above expression as $x_4(x_3' \oplus x_1) \oplus x_2$. After this, we can implement it using the expression:

$$CNOT(x_3', x_1).Toff(x_4, x_1, x_2).CNOT(x_3', x_1)$$

In the above implementation, we consider $x_1$ as an intermediate target qubit using $CNOT(x_3', x_1)$. After implementing the expression, we repeat it to remove its effect on the qubit $x_1$.

Alternatively, we can achieve the above expression through a swapping process. For example, Let's revisit Figure 2.d as presented in Figure 2.e. The cells of PI$_1$ in Figure 2.d can be extended to the neighboring cells with the value of 0 (shown as striped squares) to form $x_1'.x_3'$ and then can be swapped with the cells adjacent to PI$_2$ ($x_1.x_3'$) using a $CNOT(x_3', x_1)$ where the control qubit is equal to the common term ($x_1'.x_3' \cap x_1.x_3' = x_3'$) and the target qubit is equal to the non-common term ($x_1'.x_3' \triangle x_1.x_3' = x_1$). Next, to remove the 1's of the 4-cell PI $x_1.x_4$, $CNOT((x_1, x_4), x_2) = Toff(x_1, x_4, x_2)$ is applied. Finally, the synthesized circuit implements the following gates, which are the same as before, but we can achieve them using fewer steps.

$$CNOT(x_3', x_1).Toff(x_1, x_4, x_2).CNOT(x_3', x_1).$$

As a result, in our proposed method, we consider this advantage for some templates, which we will mention in a related section.

The third technique for simplification is employed by the groups containing both 0 and 1 values to generate larger PIs. Consider a cell of the K-map with a 1 value (i.e., one of the true minterms of the $v_i$ function). If inverting the main variable of the minterm of this cell results in a $v_i = 0$, then we refer to the group containing these two complementary cells as a Special Pair. By applying a CNOT gate with the product term of the special pair as the control qubit and the main variable as the target qubit, the $v_i$ function remains unchanged. When multiple special pairs are adjacent to each other, they create a Special Prime Implicant (SPI). Combining these SPIs with the PIs of the K-map can generate larger PIs, which have the same effect as the initial PI when considered as control qubits of the above CNOT gate. This can lead to a reduction in the number of control qubits.

For example, let's consider Figure 2.f, which shows a K-map for $v_1$ with an SPI and a PI. The SPI includes two cells with 1 value, which have indices 0000 and 1100, and two cells with 0 value, which have indices 1000 and 0100. Clearly, the value of the function for the indices (0000, 1000) is the complement of that for (1100, 0100) with respect to the main variable. These two SPs form an SPI. To generate a larger PI, we can share the adjacent cells with a minterm $x_3'.x_4$ between SPI and PI. Then, applying $CNOT(x_3', x_1).CNOT(x_4, x_1)$ leads to removing 1's of the initial PI, while the SPI part remains intact.

### 4. The proposed synthesis method

In this section, we explain the proposed transformation-based reversible synthesis method. Our objective in this method is to transform a permutation matrix (reversible matrix) into an identity matrix by finding and applying a specific sequence of equivalent matrices of gates from the NCT library. After applying these equivalent matrices, we can obtain the equivalent circuit of the initial permutation matrix by using the conjugate transpose of the applied matrices of the gates. We do this using the concept of a global view, considering all control qubits associated with CNOTs, rather than analyzing gates locally. This is achieved using K-maps or QM on each input variable. The use of QM aims to displace and group 1's of the $v_i$ functions together in larger squares to simplify and remove those 1's, resulting in fewer control qubits for controlled gates. Moreover, this method does not require ancilla qubits and the outputs are generated directly on the input qubits.

In the rest of the paper, some concepts and definitions are explained by K-maps due to its visual characteristics which can help readers understand them easier. Both K-map and QM methods are based on similar concepts but as the QM method is a more systematic approach, our algorithm is based on QM. However, due to the need for some modifications to be adapted for reversible logic, it is referred to as Modified QM (MQM).

The MQM method described in this paper differs from the traditional one. In the MQM, we aim to remove 1's (minterms) of $v_i$. The reason is that each MQM is on the XOR plane of input-output, and each 1 in the vector $v_i$ represents a difference between the input and output. By eliminating them, we can achieve our goal, which is the identity matrix. By doing this, each input variable is transformed into the output. This process is repeated for all input variables, resulting in the initial transformation matrix becoming the identity matrix. The second difference is that swapping operations are allowed on minterms by applying appropriate gates. Thirdly, each minterm can belong to only one implicant or an odd number of them.

Moreover, in Section 4, we will introduce a method that, when combined with the proposed method, can be used for reversible circuits with a large number of inputs. This approach will facilitate the exploration of the state space of the synthesis problem to achieve low-cost solutions in large reversible circuits and to make the synthesis approach scalable.

In the following, we will start by defining some definitions and conventions.

**Definition 1.** True/false minterms of a function *f* are those minterms for which $f = 1/f = 0$.

**Note**: In the remainder of the paper, the term minterm refers to a true minterm.

**Convention 1.** When two adjacent implicants are combined to create a larger implicant, if the Hamming distance of the two groups is one, the different bit is denoted by an "*X*" to indicate that the group is independent of the corresponding variable. For example:

$$\begin{Bmatrix} 0\ 1\ 1 \\ 1\ 1\ 1 \end{Bmatrix} \Rightarrow X\ 1\ 1$$

**Convention 2.** The inputs of a truth table are segmented in pairs from right to left according to their order in the list of variables. For example, for a seven-variable function $v(x_1 x_2 x_3 x_4 x_5 x_6 x_7)$, the variables are grouped as $x_1 x_2, x_3 x_4, x_5 x_6, x_7$. By fixing each variable pair and changing the others, rows, columns, and pages are created in higher dimensions. In fact, this segmentation is used to determine the direction of implicants/prime implicants of the function $v_i$ in the K-map and MQM table. (See Definition 2 for the definition of direction of minterms and implicants).

In each K-map/MQM table, the input variable $x_i$ associated with the function $v_i$ is named as the main variable. The objective in this paper is to eliminate the 1's (minterms) from each function $v_i$ and the main variable can be used to eliminate 1's of the vector $v_i$ when selected as the target qubit of CNOT gates. $x_i$ can be inside or outside the K-map/MQM table.

According to the above conventions, the following definitions are presented:

**Definition 2.** Two minterms are in the *same direction along a segment $s_k$* if their indices differ by only one segment $s_k$ and all other segments are the same. Likewise, two implicants are in the same direction if their indices have *X* values in the same segment.

For example, in Fig. 3.a, the two minterms $01,00$ and $10,00$ are in the same direction for their different second segment[§] (*i.e.*, column direction) while the minterm $11,11$ is not in the same direction as any of the other two minterms. In Fig. 3.b, the implicants $00,0X$ and $11,X1$ are in the same direction along with the first segment because they have $X$ in the same segment (*i.e.*, row direction).

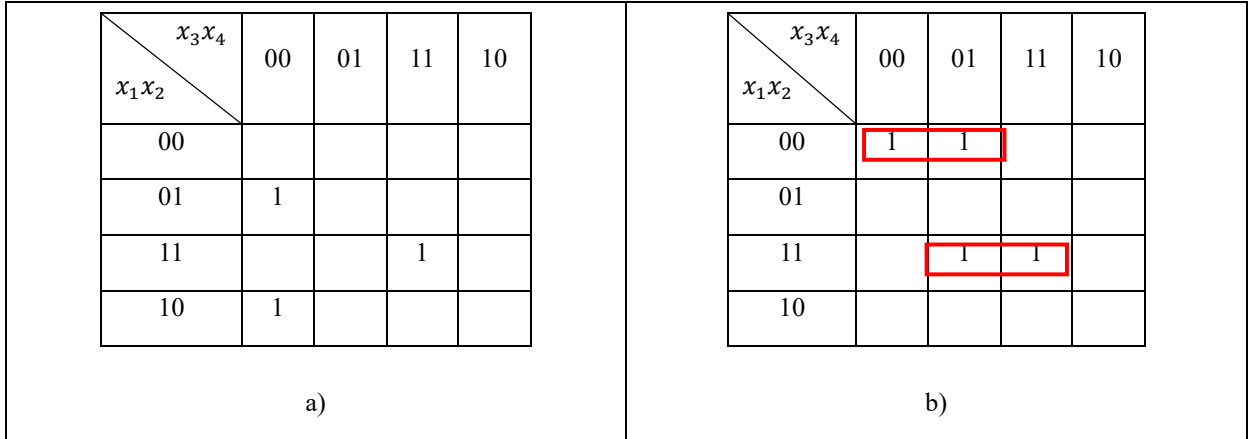

Figure 3: Illustration of Definition 2 on K-map.

**Definition 3**. Let's consider two implicants of the function $v_i$. To form larger implicants, they must be adjacent. Two implicants are *adjacent along a segment $s_k$* if they are different only in one bit in one segment ($s_k$) while all other segments are the same, resulting in a Hamming distance of a power of 2.

For example, consider two minterms with indices $(00,11)$ and $(00,10)$ in the K-map of Figure 4.a. These two minterms have a Hamming distance of $(00, 01)$, i.e., they are different in only one bit in their first segment, which makes them adjacent in that segment (Equivalent minterm is $00,0X$).

**Definition 4**. For each function $v_i$, implicants of the *same size* are defined as those with an equal number of Xs in their indices.

**Definition 5.** In a K-map, an implicant is called a *row* or a *column* if it has the value XX in at least one segment. For example, Figure 4.a shows an implicant with the index $00, XX$, representing a row, while Figure 4.b depicts an implicant with the index $XX, 00$, representing a column.

---

[§] Remember that segments are numbered from right to left.

Figure 4: Illustration of a row or column in Definition 4.

In the following, we will provide a detailed explanation of the proposed method based on the above conventions and definitions. Figure 5 shows the pseudocode for the proposed synthesis approach. An *n*-variable reversible function F = {$f_1$, ..., $f_n$} is given to the algorithm as a truth table or permutation matrix. The algorithm generates a sequence of $C^m NOT$ and NOT gates.

| | Input: reversible functions with *n* variables $x_1$, ..., $x_n$ and *n* outputs $f_1$, ..., $f_n$. Output: A reversible circuit (G) to implement $f_1$, ..., $f_n$ using NOT and $C^m$NOT gates |
|---|---|
| 1 | For each variable $x_i$ { |
| 2 |    Find $v_i = x_i \oplus f_i$. |
| 3 |    Generate $V_i$ from $v_i$ and add it to list M if $v_i \neq 0$. } |
| 4 | Sort M in ascending order with respect to the number of members of $V_i$. |
| 5 | While (M ≠ ∅) { |
| 6 |    Select the first $V_i$ from M. |
| 7 |    Remove 1's of $V_i$ by Evaluation function ($V_i$) |
| 8 |    Add appropriate gates to G. |
| 9 |    Remove $V_i$ from M. |
| 10 |    For each vector $V_j$ in M |
| 11 |       Update $V_j$ according to the applied gates; } |
| 12 | Return G. |

Figure 5: Pseudocode for the proposed synthesis approach.

|    | Input: the vector $V_i$ |
|----|---|
|    | Output: A sequence of gates (R) |
|    | Evaluation function ($V_i$) |
|    | { |
|    | /* Apply Step 1 of the algorithm. */ |
|    | { |
| 1  | Map each $V_i$ on $MQM_{x_i}$ and obtain PIs and SPIs. |
| 2  | Sort PIs and then SPIs according to their cost in ascending order |
| 3  | If there is a hamming distance >1 for specified PIs/SPIs, apply Swapping step to ($V_i$); |
| 4  | Add appropriate gates to R. |
| 5  | create larger PI for adjacent PIs with the same size |
| 6  | Remove 1's of the PI by applying CNOT on the main variable and add appropriate gates to R}. |
| 7  | while ($V_i \neq \emptyset$) |
|    | { |
|    | /* Apply Step 2 of the algorithm: */ |
| 8  | Obtain PIs using the QM algorithm |
| 9  | Check the PI templates to create larger PIs using common cells |
| 10 | If hamming distance > 1 for specified PIs, apply Swapping step ($V_i$) and add appropriate gates to R; |
| 11 | Create the larger PI. |
| 12 | Remove 1's of PI by applying CNOT on the main variable and add appropriate gates to R.} |
| 13 | Return R.} |

Figure 6: Pseudocode for evaluation function.

In the following, each line of the pseudocode in Figure 5 is explained in more detail.

Consider a truth table of an *n*-variable reversible function with input variables $(x_1, x_2, ..., x_n)$ and output variables $(f_1, f_2, ..., f_n)$. For each input-output variable pair $((x_i, f_i), i = 1,2, ..., n)$ in the truth table, a function $v_i = XOR(x_i, f_i)$ is generated (Line 2). Then, the indices of the elements of $v_i$ with a value of one (true minterms) are stored in a vector denoted as $([V_i], i = 1,2,3,4, ..., n)$. If there are true minterms in $v_i$ (i.e., $V_i \neq \emptyset$), it is added to the list of non-zero functions M (Line 3). Then, the vectors $V_i$ are prioritized according to the number of their true minterms (Line 4). Afterward, the vector $V_i$ with the fewest number of elements is selected first (Line 6).

In the Lines 7-11, each vector $V_i$ is selected, and the set of gates that can make the input qubit $q_i$ equal to $f_i$ is generated. To this end, all elements in the vector $V_i$ should be changed to 0. This is done in two phases by evaluation (Line 7 in Figure 5) and swapping (Lines 3 and 9 in Figure 6) functions. Consider the pseudocode for the evaluation function in Figure 6. In the first step, the vector $V_i$ is mapped to a corresponding MQM table named according to its main variable $(x_i)$ as $MQM_{x_i}$ (e.g., $MQM_{x_1}$ for the table where the main variable is $x_1$) (Line 1 in Figure 6). Next, the aim is to create larger implicants by identifying proper implicants, making them adjacent via the swapping function and subsequently eliminating all members of the vector $V_i$ using $C^m NOT$ gates. The algorithm first executes the evaluation phase and then the swapping phase within a while loop. The loop continues as long as there is any member in $V_i$. In fact, the two phases, evaluation and swapping, generate a sequence of gates that are applied to the qubit $q_i$ to remove all members of $V_i$ and therefore produce the target values of $f_i$.

Finally, operations applied in two phases are stored in their proper order (Line 8 in Figure 5). As $V_i = \emptyset$ now, it is removed from the list M (Line 9). The $C^m NOT$ gates applied in the swapping phase may change the truth table of unvisited qubits. As a result, the unvisited $V_i$ vectors in M need to be updated for the applied gates (Lines 10-11).

After all the qubits are processed (i.e., $V_i = \emptyset \; \forall \, i \in [1, \ldots, n]$), the post-processing simplification step is applied. This step checks the final ESOP expression, removes the redundant terms, and also factorizes the common literals in $V_i$s. Then, it considers the same gate for the same terms in different $V_i$s. Finally, the sequence of gates to obtain the synthesized circuit is returned.

The evaluation and swapping phases are explained in the following section.

### 4.1 Evaluation phase

In this phase, we aim to remove the 1's (true minterms) of the function $v_i$ (or members of the vector $V_i$) by applying $C^m NOT$ gates on the qubit $q_i$ of the main variable ($x_i$). To achieve this, we first need to identify implicants and, if possible, combine them into larger implicants that lead to complementing the true minterms in the direction of the main variable ($x_i$). Then, by applying the appropriate CNOT gate with the main qubit $q_i$ as the target qubit, the ones of $V_i$ are eliminated. This is done using the MQM method.

The evaluation phase is performed in two steps. In the first step, it is assumed that the main variable is included in the list of variables of the MQM table, and some of the minterms of $v_i$ are covered by PIs and removed. In the second step, the uncovered minterms of the main variable are covered and removed. In addition, the case where the main variable is not in the MQM table will also be handled.

**Step 1**: If the main variable is present in the indices of the true minterms, there will be a new degree of freedom for further simplification. The following steps check this case:

1- Consider the equivalent index corresponding to each element $V_i(j)$ of the vector $V_i$. The main variable of the index is complemented to produce $\hat{V}_i(j)$. Then, $V_i(j)$ is scored using the following evaluation function F and the scores are stored in a new vector $V2_i$. This vector has two columns. The first column contains the index $V_i(j)$ and the second column contains its score. The score of the $j$th element of $V_i$ is calculated as follows:

$$F(j) = \begin{cases} 0 & if \; (v(V_j), v(\hat{V}_j)) = (1,0) \\ -1 & if \; (v(V_j), v(\hat{V}_j)) = (1,1) \end{cases}$$

We call $\hat{V}_i(j)$ the complement of $V_i(j)$ in the direction of the main variable. If both complementary indices exist in the vector $V_i$, only the first one is kept in $V2_i$.

If $F(j) = 0$, it indicates that two adjacent cells in the K-map with complementary indices in the direction of the main variable have values 1 and 0. This function can help identify and utilize the SPIs defined in the previous section. If $F(j) = -1$, it indicates that two cells in the K-map with complementary indices in the direction of the main variable have a value of 1, and therefore, swapping them can eliminate two ones (true minterms). This concept can be restated as Lemma 1 using quantum gates.

**Lemma 1**: Consider the vector $V_i$ for the function $v_i$ with the main variable $x_i$. If $F(j) = -1$, then applying a CNOT gate where its control qubit covers two indices of $V_i(j)$ and $\hat{V}_i(j)$ and its target qubit is the qubit $q_i$ (corresponding to the input variable $x_i$), removes the two specified minterms of the function $v_i$.

Proof: The vector $V_i$ contains the indices of the rows of the truth table where there is a difference between the input $x_i$ and the output $f_i$. By applying a CNOT gate where the control qubits cover the specified indices and the target qubit is equal to the main variable $x_i$, the two indices are swapped in the direction of the main variable. This operation removes the difference between the input and output columns for these indices in the truth table, making the input equal to the output for the corresponding rows of the truth table.

2- In this step, the elements of each pair of indices in the vector $V2_i$ are compared to each other ignoring their main variable's qubit and Line 2 of the QM algorithm is applied to the remaining qubits. Therefore, The Hamming distance of the remaining qubits of each index pair is calculated. If a pair differs by only one qubit within a segment, their corresponding row indices, the sum of their scores, and their indices are recorded in a new vector $V3_i$. As stated in Definition 3, the index generated from the two elements is created by replacing their differing qubit with an 'X', representing the difference between the indices of the two elements. Mark any two indices that can be combined. This step is repeated to form PI/SPIs and new columns ($V4_i$, $V5_i$, etc.) are formed until no pair of indices with one different qubit value is found.
3- Sort the PI in ascending order. Then Step 4 of the QM method is applied for the obtained PI.
4- The columns of the above table are checked and the PI with the most negative score is selected. Each minterm must be covered by one PI. Especially, if a minterm is covered by an even number of other PIs, those PIs are removed. If it is covered by an odd number of other PIs, and those PIs have minterms that are not covered by other PIs, they are also kept. Moreover, among the PIs with identical scores, the one aligned with the previously chosen PI is selected.
5- Remove the rows related to the selected PI. Repeat the above step in the reduced table for other PIs with negative costs until no PIs with negative costs are left.
6- Repeat Steps 2 to 5 for the new PI/SPIs until all the PIs are obtained, and no PI is combined with another.
7- To simplify the circuit, larger PIs are formed from the obtained PI/SPIs of the previous step without common minterms as follows:

    (a) The PIs are sorted in ascending order. This means that we start by selecting the PIs with the most negative scores. Then, we add PIs with the same or smaller negative score and with the same size and direction as the previous PIs, one by one. In the final stage, we consider PIs with a score of zero (referred to as SPIs in the previous section) and try to use them together to form a larger PI. The number of selected minterms must ultimately be a power of 2. It is important to note that the selected PIs must have 'X' symbol within the same position of the same segment, indicating the same row or column direction. As stated in Definition 3, to create a larger PI using two given PIs, they need to be placed adjacent to each other. This means that the Hamming distance between the two PIs should be 1 for one segment of their indices, and 0 for other segments. If the Hamming distance is more than one, a swapping phase repositions the PIs closer together (see swapping phase section for details).

    (b) To remove elements of the vector $V_i$, Eqs. 1 to 3 are used. In this case, the selected PI can be written down as the following ESOP expression:
$$PI \oplus x_i$$
Then, the above expression is saved and transformed into $CNOT(PI, x_i)$. This operation can either be applied as a right-hand multiplication to the permutation matrix or directly to the input states of the truth table.

For general cases when there are many PIs, such as PI1, PI2, …, the following CNOT gate can be applied to the input:
$$(PI1 \oplus PI2 \oplus PI3) \oplus x_i$$
$$=> CNOT(PI1, x_i) \times CNOT(PI2, x_i) \times CNOT(PI3, x_i)$$

Then for the uncovered 1's, the procedure in Step 2 is applied.

**Step 2**: If the indices of the true minterms do not include the main variable or if the vector $V_i$ has some elements that have not been covered in Step 1, they are processed in this step.

First, the traditional QM algorithm is applied to the uncovered 1's to remove the existing minterms. In this process, each minterm must be covered by a single PI, the same as before. If a minterm is covered by an even number of other PIs, those PIs are removed. If it is covered by an odd number of other PIs, and those PIs have minterms that are not covered by other PIs, they are kept. Then, PIs are compared against a set of predefined templates as follows. If a match is found, the corresponding ESOP expression is saved and the corresponding CNOT gate can either be applied as a right-hand multiplication to the permutation matrix or directly to the input states of the truth table.

> **Matching Templates using common minterms:** In a K-map, if two PIs can potentially constitute a larger PI using additional minterms as stated in the following templates, then the PIs are expanded and combined together to create a larger one. In fact, these additional minterms are used twice as common minterms of two PIs (i.e. $1 \oplus 1 = 0$). Therefore, it does not affect the function $v_i$. Finally, by considering Eqs. (1) to (3), the corresponding CNOT gates can either be applied as a right-hand multiplication to the permutation matrix or directly to the input states of the truth table.
>
> To distinguish two PIs with the mentioned characteristics, the PIs must be adjacent by sharing minterms and meet one of the following conditions.

1) Consider two equal-sized PIs, A and B, represented by two indices of segmented qubits:
   A = ..., $XX$, ..., $\alpha\beta$, ...
   B = ..., $XX$, ..., $\bar{\alpha}\bar{\beta}$, ...
   where the positions of the respective segments are the same for both strings and the bar on a symbol represents logical inversion[**]. In other words, $XX$ represents don't cares for the segment $x_j x_{j+1}$ in both strings and both $\alpha\beta$ and $\bar{\alpha}\bar{\beta}$ correspond to the segment $x_k x_{k+1}$.[††] In this template, the remaining qubits in the two strings of A and B, shown by dots, are identical. The identical qubits can be either 0, 1 or X. These PIs can generate two larger PIs, namely A' and B', by combining their minterms with the common minterms with indices $A_{ext}$ = ..., $XX$, ..., $\alpha\bar{\beta}$, ...:
   A' = A ◇ $A_{ext}$ = ..., $XX$, ..., $\alpha X$, ...

---

[**] In all templates, the order of segments can be reversed. For example, the segment $\alpha\beta$ and $\bar{\alpha}\bar{\beta}$ can be on the left-hand side of $XX$.
[††] This notation for the orders of the qubits is also used for other templates.

B' = B ◇ A_ext = ..., $XX$, ..., $X\bar{\beta}$, ...

where the symbol ◇ is used for combining two PIs into one larger PI. Alternatively, A and B could be extended by combining them with B_ext = ..., $XX$, ..., $\bar{\alpha}\beta$, ...:

A' = A ◇ B_ext = ..., $XX$, ..., $X\beta$, ...
B' = B ◇ B_ext = ..., $XX$, ..., $\bar{\alpha}X$, ...

Then, for the first case, the true minterms of A and B are removed by a CNOT gate with $(x_k^\alpha \oplus x_{k+1}^{\bar{\beta}})$ as its control qubit and the main variable as its target qubit where $x_m^n$ is defined as:

$$x_m^n = \begin{cases} x_m & \text{if } n = 1 \\ \overline{x_m} & \text{if } n = 0 \end{cases}$$

This gate will not affect the value of the function in the groups consisting of indices A_ext or B_ext because they are covered twice. Figure 2.c depicts the K-map representation of this template for a four-variable function where A = $XX$,00 and B = $XX$,11. The ESOP expression and the CNOT gate related to the template are represented as follows:

$$(x_k^\alpha \oplus x_{k+1}^\beta) \oplus x_i \rightarrow CNOT(x_k^\alpha \oplus x_{k+1}^\beta, x_i)$$

where $x_i$ is the main variable. According to Eqs. 1 to 3, the above expression can be stated in two ways:

$$CNOT(x_{k+1}^\beta, x_k^\alpha).CNOT(x_k^\alpha, x_i).CNOT(x_{k+1}^\beta, x_k^\alpha)$$

or

$$CNOT(x_k^\alpha, x_i), CNOT(x_{k+1}^\beta, x_i)$$

2) In this template, there are two PIs, A and B, with segments represented by

A = ..., $0X$, ..., $\alpha_1\beta_1$, ...
B = ..., $1X$, ..., $\alpha_2\beta_2$, ...

where the Hamming distance of $\alpha_1\beta_1$ and $\alpha_2\beta_2$ equals to 1. One of the PIs (e.g., A) is selected as the base PI. A set of common minterms is formed by replacing the segment $1X$ in B with the corresponding segment $0X$ in A. This creates a new PI (A') that is adjacent to A and is in the same direction with B:

A_ext = ..., $1X$, ..., $\alpha_1\beta_1$, ...

Then, A and B can be extended as:

A' = A ◇ A_ext = ..., $XX$, ..., $\alpha_1\beta_1$, ...
B' = B ◇ A_ext = ..., $1X$, ..., $X\beta_1$, ... if $\alpha_1 \neq \alpha_2$ and $\beta_1 = \beta_2$

$$= \ldots, 1X, \ldots, \alpha_1 X, \ldots \text{ if } \alpha_1 = \alpha_2 \text{ and } \beta_1 \neq \beta_2$$

In this case, in order to remove the true minterms of $v_i$ in A and B, the control qubits of the CNOT gate will be provided by the expression $(x_k^{\alpha_1} x_{k+1}^{\beta_1} \oplus x_j x_{k+1}^{\beta_1})$ if $\alpha_1 \neq \alpha_2$ or by $(x_k^{\alpha_1} x_{k+1}^{\beta_1} \oplus x_j x_k^{\alpha_1})$, otherwise. The target qubit will be the main variable. The values of $v_i$ for the cells of $A_{ext}$ are left intact as they are covered twice. Figure 2.d shows the K-map representation of this template for a four-variable function where A = 0X,01 and B = 1X,X1.
Another version of this template is as follows:
A = ..., X0, ..., $\alpha_1 \beta_1$, ...
B = ..., X1, ..., $\alpha_2 \beta_2$, ...
A similar discussion can also be derived for this template. In the following cases, the order of qubits for a segment like X0 can be 0X for being used as a template as long as the X for A is at the same position as the X in B.

3) Consider two PIs, A and B, with A being twice the size of B:
A = ..., XX, ..., $\alpha_1 \beta_1$, ...
B = ..., 1X, ..., $\alpha_2 \beta_2$, ...
or
B = ..., 0X, ..., $\alpha_2 \beta_2$, ...
where the Hamming distance of $\alpha_1 \beta_1$ and $\alpha_2 \beta_2$ equals to 1.
Let B = ..., 0X, ..., $\alpha_2 \beta_2$, ....
There is a group $B_{ext}$ neighboring B which can be used as common minterms:
$B_{ext}$ = ...,1X, ..., $\alpha_2 \beta_2$, ...
Then B can be expanded as:
$B_2 = B \diamond B_{ext} = \ldots, XX, \ldots, \alpha_2 \beta_2, \ldots$
Now, $B_2$ and A can generate a PI larger than A (namely, A') while $B_{ext}$ is used as the other PI to neutralize the additional cells covered by A'. As a result, the new PIs are as follows:
$$A' = A \diamond B_2 = \ldots, XX, \ldots, X\beta_1, \ldots \text{ if } \alpha_1 \neq \alpha_2 \text{ and } \beta_1 = \beta_2$$
$$= \ldots, XX, \ldots, \alpha_1 X, \ldots \text{ if } \alpha_1 = \alpha_2 \text{ and } \beta_1 \neq \beta_2$$
B' = $B_{ext}$

Therefore, The control qubit of a CNOT gate is provided by the expression $(x_{k+1}^{\beta_1} \oplus x_j x_k^{\alpha_2} x_{k+1}^{\beta_2})$ if $\alpha_1 \neq \alpha_2$ or by $(x_k^{\alpha_1} \oplus x_j x_k^{\alpha_2} x_{k+1}^{\beta_2})$, otherwise. The target will be the qubit corresponding to the main variable. Figure 7.a shows the K-map representation of this template for a four-variable function where A = *XX,01* and B = *1X,11*. In this example, B' = $B_{ext}$ = 0X,11, $B_2$ = XX,11, and A' = A $\diamond$ $B_2$ = XX,X1.

For the second case where B = ..., 0X, ..., $\alpha_2\beta_2$, ..., the CNOT gates can be derived similarly.

4) Consider two PIs, A and B, where A is twice the size of B:
A = ..., $\alpha_1 X$, ..., $\beta_1 X$, ...
B = ..., $\alpha_2 X$, ..., $\beta_2 \gamma_2$, ...
and the Hamming distance of $\alpha_1\beta_1$ and $\alpha_2\beta_2$ equals to 1. Applying a NOT gate to the variable corresponding to the position of $\gamma_2$ (i.e., $q_{j+1}$), generates a group as common minterm:
$B_{ext}$ = ..., $\alpha_2 X$, ..., $\beta_2 \overline{\gamma_2}$, ....
Then, B can be extended as $B_2$ to generate a group to be combined with A and produce a larger PI:
$B_2$ = ..., $\alpha_2 X$, ..., $\beta_2 X$, ...
A' = A $\diamond$ $B_2$ = ..., $XX$, ..., $\beta_1 X$, ... if $\alpha_1 \neq \alpha_2$ and $\beta_1 = \beta_2$
= ..., $\alpha_1 X$, ..., $XX$, ... if $\alpha_1 = \alpha_2$ and $\beta_1 \neq \beta_2$
B' = $B_{ext}$

The control qubits of a CNOT gate will be provided by the expression $(x_k^{\beta_1} \oplus x_j^{\alpha_2} x_k^{\beta_2} x_{k+1}^{\overline{\gamma_2}})$ if $\alpha_1 \neq \alpha_2$ or by $(x_j^{\alpha_1} \oplus x_j^{\alpha_2} x_k^{\beta_2} x_{k+1}^{\overline{\gamma_2}})$, otherwise. The target will be the qubit $q_i$ of the main variable.
An example of this template for a four-variable function is shown in Figure 7.b where A = 0X,0X and B = 00,1X.

5) Consider two equal-sized PIs, A and B:
A = ..., $\alpha_1 X$, ..., $\beta_1 X$, ...
B = ..., $\overline{\alpha_1} X$, ..., $\overline{\beta_1} X$, ...
Then, either of the following extensions:
$A_{ext,1}$ = ..., $\overline{\alpha_1} X$, ..., $\beta_1 X$, ...
or
$A_{ext,2}$ = ..., $\alpha_1 X$, ..., $\overline{\beta_1} X$, ...
can be used as common minterms to generate two larger PIs, namely A' and B':
A' = A $\diamond$ $A_{ext,1}$ = ..., $XX$, ..., $\beta_1 X$, ...
B' = B $\diamond$ $A_{ext,1}$ = ..., $\overline{\alpha_1} X$, ..., $XX$, ...
or
A' = A $\diamond$ $A_{ext,2}$ = ..., $\alpha_1 X$, ..., $XX$, ...
B' = B $\diamond$ $A_{ext,2}$ = ..., $XX$, ..., $\overline{\beta_1} X$, ...

The CNOT gate has a control qubit as $(x_k^{\beta_1} \oplus x_j^{\overline{\alpha_1}})$ or $\left(x_j^{\alpha_1} \oplus x_k^{\overline{\beta_1}}\right)$ and a target qubit as the main variable. Figure 7.c depicts an example of this case.

6) Consider three PIs, $A$, $B$ and $C$:
   $A = \ldots, \alpha X, \ldots, \beta_1 \gamma_1, \ldots$
   $B = \ldots, \delta \zeta, \ldots, \beta_2 \gamma_2, \ldots$
   $C = \ldots, \delta \overline{\zeta}, \ldots, \beta_2 \gamma_2, \ldots$
   where the Hamming distance of $\beta_1 \gamma_1$ and $\beta_2 \gamma_2$ equals to 1. Therefore, they can be considered adjacent with respect to the corresponding segments, and a common implicant $A_{ext}$ can be identified for them as the following:
   $A_{ext} = \ldots, \delta \overline{\zeta}, \ldots, \beta_2 \gamma_2, \ldots$ if $\alpha = \delta$
   $\phantom{A_{ext} =} \ldots, \overline{\delta} \zeta, \ldots, \beta_2 \gamma_2, \ldots$ if $\alpha \neq \delta$

   This common minterm, combined with PIs B and C, forms the following extended PIs:

   $B' = B \Diamond A_{ext} = \ldots, \delta X, \ldots, \beta_2 \gamma_2, \ldots$ if $\alpha = \delta$
   $\phantom{B' = B \Diamond A_{ext} =} \ldots, X \zeta, \ldots, \beta_2 \gamma_2, \ldots$ if $\alpha \neq \delta$

   $C' = C \Diamond A_{ext} = \ldots, X \overline{\zeta}, \ldots, \beta_2 \gamma_2, \ldots$ if $\alpha = \delta$
   $\phantom{C' = C \Diamond A_{ext} =} \ldots, \overline{\delta} X, \ldots, \beta_2 \gamma_2, \ldots$ if $\alpha \neq \delta$

   If $\alpha = \delta$, then the combination of A with B' results in a larger PI as the following:
   $A'_1 = B' \Diamond A = \ldots, \delta X, \ldots, \beta_2 X, \ldots$ if $\alpha = \delta$ and $\beta_1 = \beta_2$
   $\phantom{A'_1 = B' \Diamond A =} \ldots, \delta X, \ldots, X \gamma_2, \ldots$ if $\alpha = \delta$ and $\beta_1 \neq \beta_2$
   If $\alpha \neq \delta$, then the combination of A with C' results in a larger PI as the following:
   $A'_2 = C' \Diamond A = \ldots, \alpha X, \ldots, \beta_2 X, \ldots$ if $\alpha \neq \delta$ and $\beta_1 = \beta_2$
   $\phantom{A'_2 = C' \Diamond A =} \ldots, \alpha X, \ldots, X \gamma_2, \ldots$ if $\alpha \neq \delta$ and $\beta_1 \neq \beta_2$

   The CNOT gate has a control qubit as $A'_1 \oplus C'$ or $A'_2 \oplus B'$ according to different conditions as follows:
   $(x_j^{\delta} x_k^{\beta_2} \oplus x_{j+1}^{\overline{\zeta}} x_k^{\beta_2} x_{k+1}^{\gamma_2})$ if $\alpha = \delta$ and $\beta_1 = \beta_2$
   $(x_j^{\overline{\delta}} x_k^{\beta_2} \oplus x_{j+1}^{\zeta} x_k^{\beta_2} x_{k+1}^{\gamma_2})$ if $\alpha \neq \delta$ and $\beta_1 = \beta_2$
   $(x_j^{\delta} x_{k+1}^{\gamma_2} \oplus x_{j+1}^{\overline{\zeta}} x_k^{\beta_2} x_{k+1}^{\gamma_2})$ if $\alpha = \delta$ and $\beta_1 \neq \beta_2$
   $(x_j^{\overline{\delta}} x_{k+1}^{\gamma_2} \oplus x_{j+1}^{\zeta} x_k^{\beta_2} x_{k+1}^{\gamma_2}$ if $\alpha \neq \delta$ and $\beta_1 \neq \beta_2$

where the target qubit corresponds to the main variable. Figure 7.d depicts an example of this case.

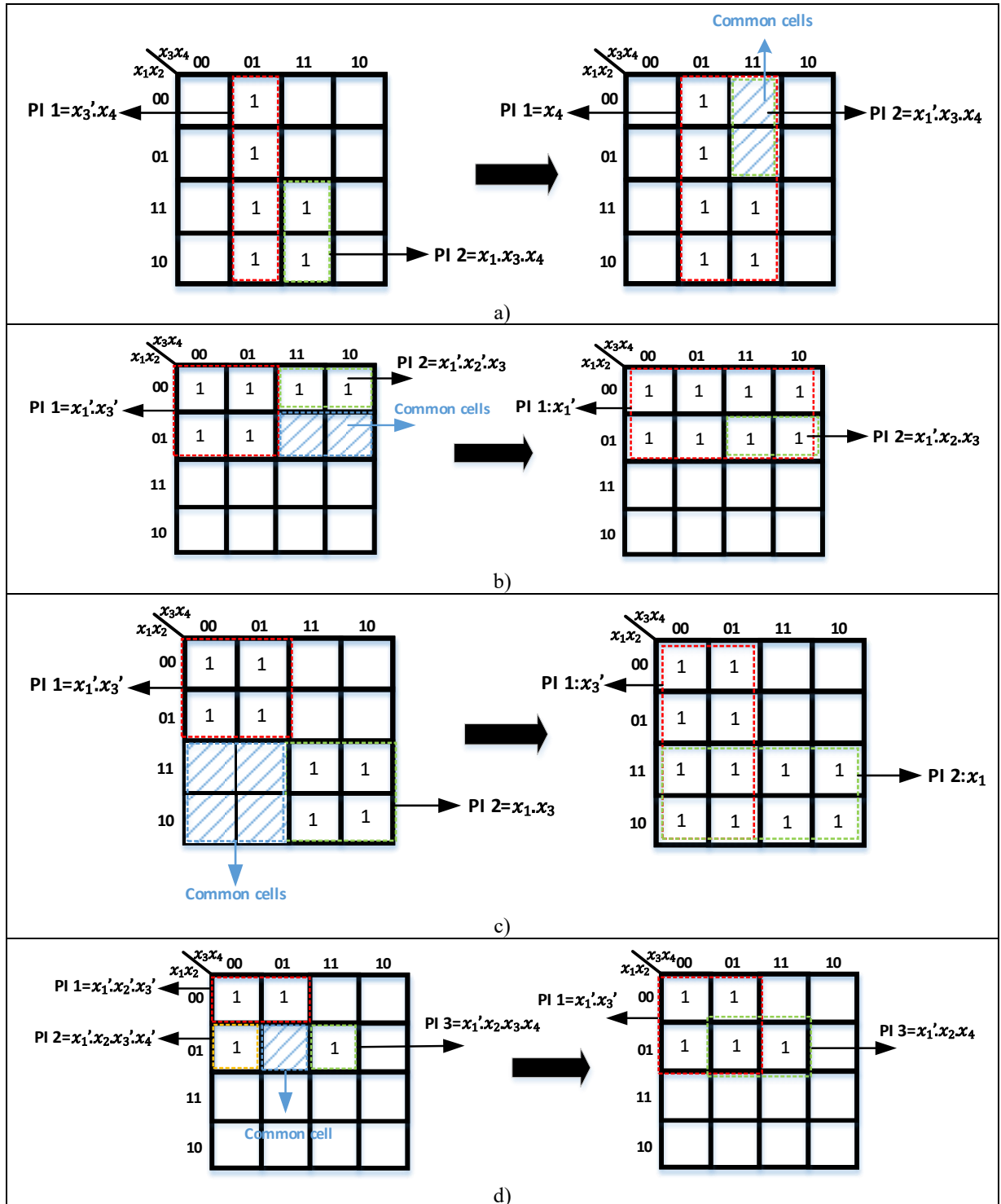

Figure 7: An example of K-map representation of some of the templates using common minterms (Line 11 from Figure 6).

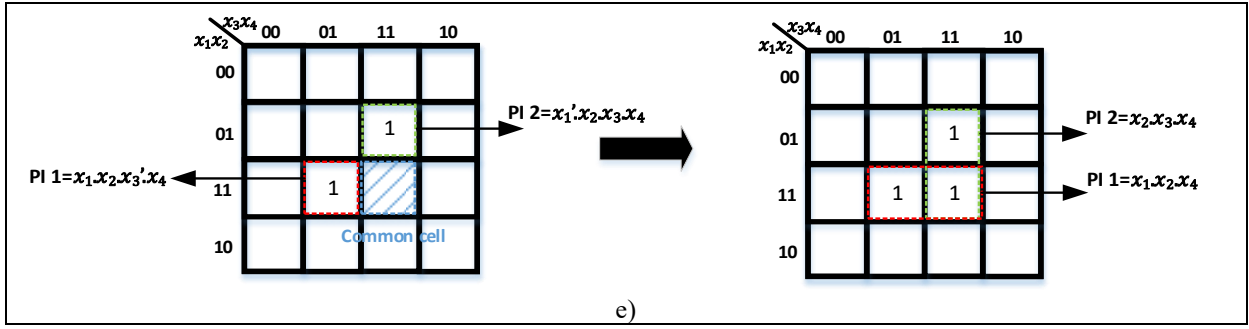
Figure 7: An example of K-map representation of some of the templates using common minterms (Line 11 from Figure 6) (continued).

7) Consider two equal-sized PIs, A and B:
   $A = \ldots, \alpha_1\beta_1, \ldots, \gamma_1\delta_1, \ldots$
   $B = \ldots, \alpha_2\beta_2, \ldots, \gamma_2\delta_2, \ldots$
   where the Hamming distance of $\gamma_1\delta_1$ and $\gamma_2\delta_2$ equals to 2. First, let $\alpha_1 = \alpha_2$ and $\beta_1 = \beta_2$. Then, a common implicant can be considered as the following:
   $A_{ext,1} = \ldots, \alpha_1\beta_1, \ldots, \gamma_1\delta_2, \ldots$
   or
   $A_{ext,2} = \ldots, \alpha_1\beta_1, \ldots, \gamma_2\delta_1, \ldots$

   The combination of A and B with the above common implicant results in larger PIs as the following:
   $A' = A_{ext,1} \Diamond A = \ldots, \alpha_1\beta_1, \ldots, \gamma_1 X, \ldots$
   $B' = A_{ext,1} \Diamond B = \ldots, \alpha_1\beta_1, \ldots, X\delta_2, \ldots$
   or
   $A' = A_{ext,2} \Diamond A = \ldots, \alpha_1\beta_1, \ldots, X\delta_1, \ldots$
   $B' = A_{ext,2} \Diamond B = \ldots, \alpha_1\beta_1, \ldots, \gamma_2 X, \ldots$
   Then the CNOT gate has a control qubit as $A' \oplus B'$ as the following:
   $$x_j^{\alpha_1} x_{j+1}^{\beta_1} x_k^{\gamma_1} \oplus x_j^{\alpha_1} x_{j+1}^{\beta_1} x_{k+1}^{\delta_2}$$
   or
   $$x_j^{\alpha_1} x_{j+1}^{\beta_1} x_{k+1}^{\delta_1} \oplus x_j^{\alpha_1} x_{j+1}^{\beta_1} x_k^{\gamma_2}$$
   where the target qubit is the main variable.
   If the Hamming distance is one for each of the segments, then, a common implicant can be considered as the following:
   $A_{ext,1} = \ldots, \alpha_1\beta_1, \ldots, \gamma_2\delta_2, \ldots$
   or
   $A_{ext,2} = \ldots, \alpha_2\beta_2, \ldots, \gamma_1\delta_1, \ldots$
   The combination of A and B with the above common implicants results in larger PIs as the following:
   $A' = A_{ext,1} \Diamond A = \ldots, \alpha_1\beta_1, \ldots, \gamma_1 X, \ldots$ if $\gamma_1 = \gamma_2$

$$A' = A_{ext,1} \diamond A = \ldots, \alpha_1\beta_1, \ldots, X\delta_1, \ldots \text{ if } \gamma_1 \neq \gamma_2$$

$$B' = A_{ext,1} \diamond B = \ldots, \alpha_1 X, \ldots, \gamma_2\delta_2, \ldots \text{ if } \alpha_1 = \alpha_2$$
$$B' = A_{ext,1} \diamond B = \ldots, X\beta_1, \ldots, \gamma_2\delta_2, \ldots \text{ if } \alpha_1 \neq \alpha_2$$

or

$$A' = A_{ext,2} \diamond A = \ldots, \alpha_1 X, \ldots, \gamma_1\delta_1, \ldots \text{ if } \alpha_1 = \alpha_2$$
$$A' = A_{ext,2} \diamond A = \ldots, X\beta_1, \ldots, \gamma_1\delta_1, \ldots \text{ if } \alpha_1 \neq \alpha_2$$

$$B' = A_{ext,2} \diamond B = \ldots, \alpha_2\beta_2, \ldots, \gamma_2 X, \ldots \text{ if } \gamma_1 = \gamma_2$$
$$B' = A_{ext,2} \diamond B = \ldots, \alpha_2\beta_2, \ldots, X\delta_2, \ldots \text{ if } \gamma_1 \neq \gamma_2$$

The CNOT gate has a control qubit as A' $\oplus$ B' and a target qubit as the main variable. Figure 7.e depicts an example of this case.

### 4.2 Swapping phase

The true minterms of $v_i$ function (the elements of the vector $V_i$) that need to be grouped during the evaluation phase should initially be positioned in adjacent cells in the K-map of the vector $V_i$. However, if they are not already adjacent, they can be moved to be arranged in adjacent cells using the method outlined in this phase. Moreover, for the templates discussed in the previous section, if the Hamming distance of two segments is equal to 2, the swapping phase is performed.

During the swapping phase, it is important to note that the operations performed do not affect the number of true minterms of the function $v_i$. In this context, the operations are performed by CNOT gates that have a target qubit other than the main variable of the vector $V_i$ ($x_i$). If the same segments of two PIs (at positions $j$ and $j+1$) were detected in the evaluation phase to have a Hamming distance greater than one:

$$A = \ldots, \alpha\beta, \ldots$$
$$B = \ldots, \bar{\alpha}\bar{\beta}, \ldots$$

then a CNOT gate can be applied to bring the two PIs adjacent to each other. The control qubit of this gate will be one of the two different qubits and the target qubit will be the other qubit of this segment. There are four options to apply the CNOT gate:

$\text{CNOT}(x_j^\alpha, x_{j+1}^\beta)$

$\text{CNOT}(x_j^\beta, x_{j+1}^\alpha)$

$\text{CNOT}(x_j^{\bar{\alpha}}, x_{j+1}^\beta)$

$$\text{CNOT}(x_j^{\bar{\beta}}, x_{j+1}^{\alpha})$$

Among these options, the one which is more effective in reaching the final output (i.e., leads to the least number of steps to achieve the identity matrix) is chosen.

In the above cases, if the target qubit $q_i$ is not currently in the list M (i.e., $V_i = 0$), then, the swapping gate is applied on $q_i$ to generate unwanted 1's of $v_i$. Therefore, after performing the corresponding evaluation phase, the swapping gate are repeated on the target $q_i$ so that these effects are removed. If the target qubit is in M, there is no need to add the extra swapping gate in the end, as it can be handled in the succeeding iterations of the main loop.

In the swapping phase, SPIs are also taken into account and are moved to make adjacent PIs or SPIs in the same way as explained above for PIs. This case is shown in the following example.

Example 1: Consider the truth table in Figure 1. The vector $V_i$ with the largest number of true minterms is chosen. Let this vector be $V_1$. First the proposed algorithm is applied to $V_1$. Using the F function, the scores of PIs and SPIs are calculated. This is shown in Figure 8.a where PIs and SPIs are marked with red boxes. The PIs/SPIs with negative scores, and then those with zero scores that match in size and direction are selected, and the Hamming distances of their indices are checked. Here, SPI1 and PI1 have scores of 0 and -2 with the same direction and size, respectively, whereas SPI2 and SPI3 have the same score of 0 but they differ in size. Since the Hamming distance of SPI1 and PI1 is one, a common PI can be defined to make them adjacent. Figure 8.b illustrates this, where two larger SPI and PI are formed, as shown in red and green. To align and remove minterms with the main variable, according to Template 1, the following gates are used:

$$CNOT(x_3', x_1), CNOT(x_4, x_1)$$

Figure 8.c shows the K-map resulting from applying these gates. Using Step 2 of the Evaluation procedure, the remaining minterms are grouped into PIs A and B. These PIs are in the same direction as stated in the template 6. The cell (10,00) of B is aligned with the cell (01,00) of A, but the Hamming distance of their indices is 2. The swapping procedure applies CNOT($x_1, x_2$) and swaps the last two rows of K-map to make the two PIs adjacent to each other. This results in a larger PI, as shown in red in Figure 8.e. Then another swapping procedure applies a $CNOT(x_3, x_4)$ to swap the minterm with index (11,10) closer to the common cell (11,01), forming another PI shown in green in Figure 8.e. Finally, by applying a CNOT gate, with the control qubit as the XOR of the two PIs and the target qubit as the main variable, all the ones are eliminated from the map. The resulting expression involves the application of two Toffoli gates $Toff(x_1, x_4, x_3)$ and $Toff(x_2, x_3', x_1)$, as shown in the following equation. $Toff(x_1, x_4, x_3)$ is repeated to remove its effect on the function $v_3$.

$$(x_3' x_2 \oplus x_1 x_2 x_4) \oplus x_1 \rightarrow x_2(x_3' \oplus x_1 x_4) \oplus x_1$$
$$\rightarrow Toff(x_1, x_4, x_3), Toff(x_2, x_3', x_1), Toff(x_1, x_4, x_3)$$

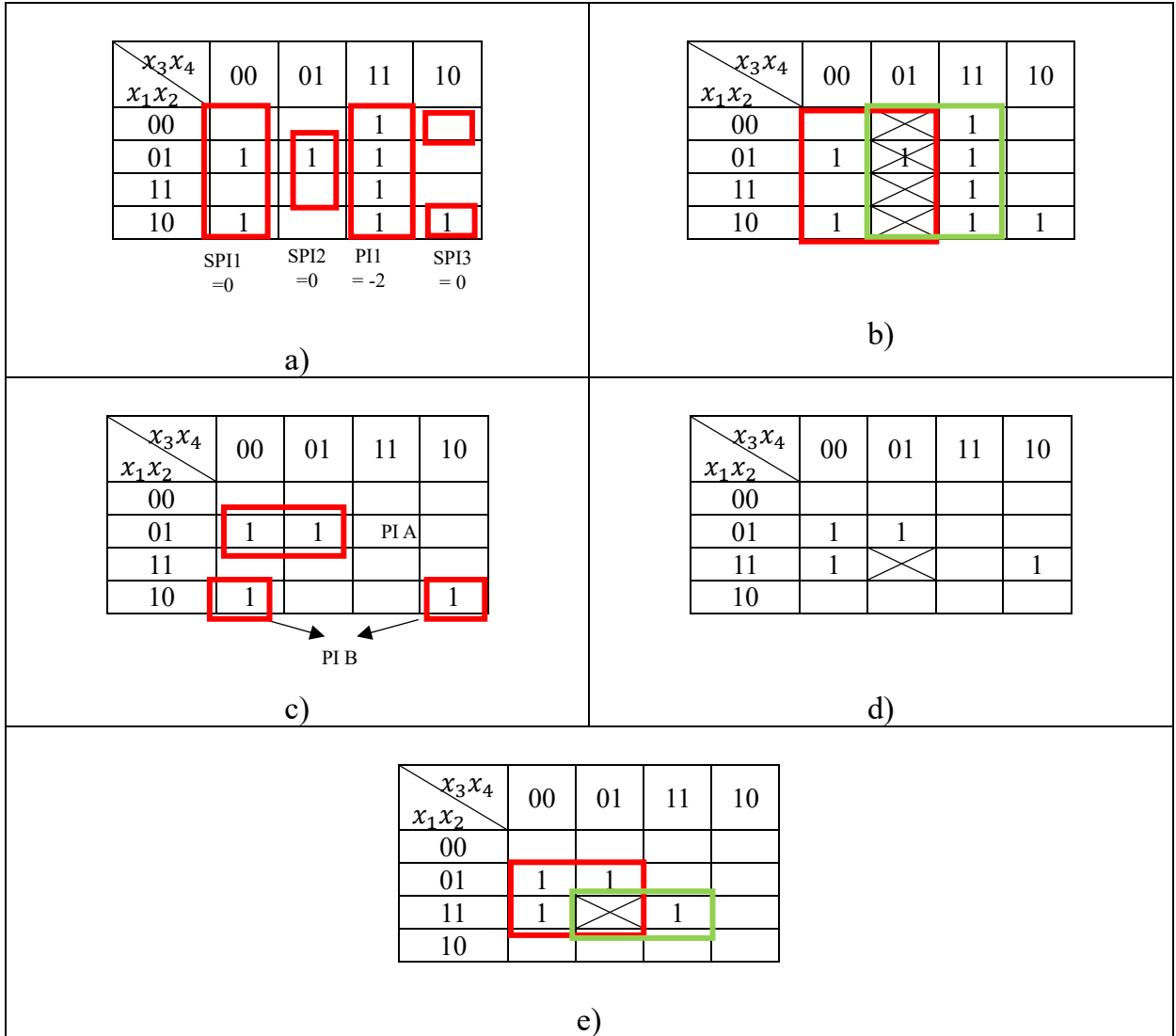

Figure 8: Illustration of K-map presented in Fig. 1 and applying the proposed synthesis method on it.

## 4.3 Post-processing simplification step

In this step, the ESOP obtained for each vector $V_i$ is investigated as the following:

1- For each vector $V_i$, if there are two common product terms, they are removed.

2- For each vector $V_i$, if there are any common literals, they are factorized.

Let $f_i$ be an $n$-variable reversible function which includes two CNOTs with $(K1 + M)$-qubits and $(K2 + M)$-qubits, respectively as shown in Figure 9.a where $x_i$ is its target qubit, $M$ is the number of common qubits, and $K1, K2$ are the number of uncommon qubits in the two gates. In this case, there are two scenarios as shown in Figures 9.b and c. If $K2 = 1$, the template shown in Figure 9.b is used. If $K2 > 1$ and $K1 > 1$, then the template shown in Figure 9.c can be used only when there is a qubit which is not used by these two gates (here named as $x_m$). Otherwise, this template is not used.

3- For each two vectors $V_i$ and $V_j$ ($i \neq j$), if there are any common literals or common product terms, they are factorized. In this case, we will consider one common CNOT gate for them, as shown in Figures 9.d-g.

Let $f_i$ and $f_j$ ($i \neq j$) be two $n$-variable reversible functions with $M$-qubit and $(K + M)$-qubit CNOT gates, respectively as shown in Figure 9.d, where $K$ and $M$ denote the number of uncommon and common qubits, respectively. If $m \geq 2$, then $f_j$ can be implemented by a single $Toff(y_{p1}, \ldots, y_{pM}, x_i)$ gate sandwiched by two $(K + 1)$-qubits CNOT gates as $CNOT(x_i, y_{r1}, \ldots, y_{r1+rK}, x_j)$ where $y_{p1}, \ldots, y_{pM}$ are the $M$ common qubits and $y_{r1}, \ldots, y_{r1+rK}$ are the $K$ uncommon qubits in the figure.

Figure 9.e illustrates a scenario where $f_i$ and $f_j$ have $M$ common qubits with uncommon qubits $K1$ and $K2$, respectively. In this case, for $m \geq 2$ we need to use one unused qubit to simplify these gates. Again, if there is a qubit which has not been used by these two gates (namely $x_m$), we can use it according to this template. Figures 9.b and 9.d illustrate templates for the cases where the common $M$ qubits in $f_i$ and $f_j$ are complementary to each other. The proofs of the templates are given as equations on the figure. In the equations, $\kappa_1, \kappa_2$ and $\mu$ stand for the product terms of the $K1$, $K2$, and $M$ qubits as shown in the figure.

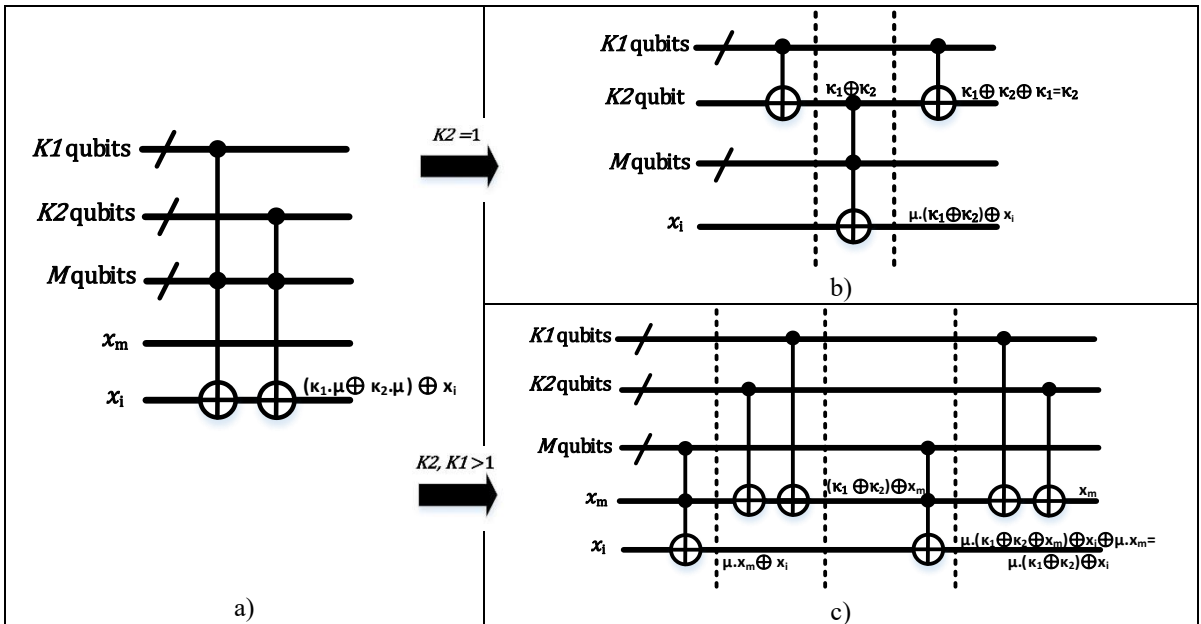

Figure 9. Factorizing common product terms with the proof of the templates.

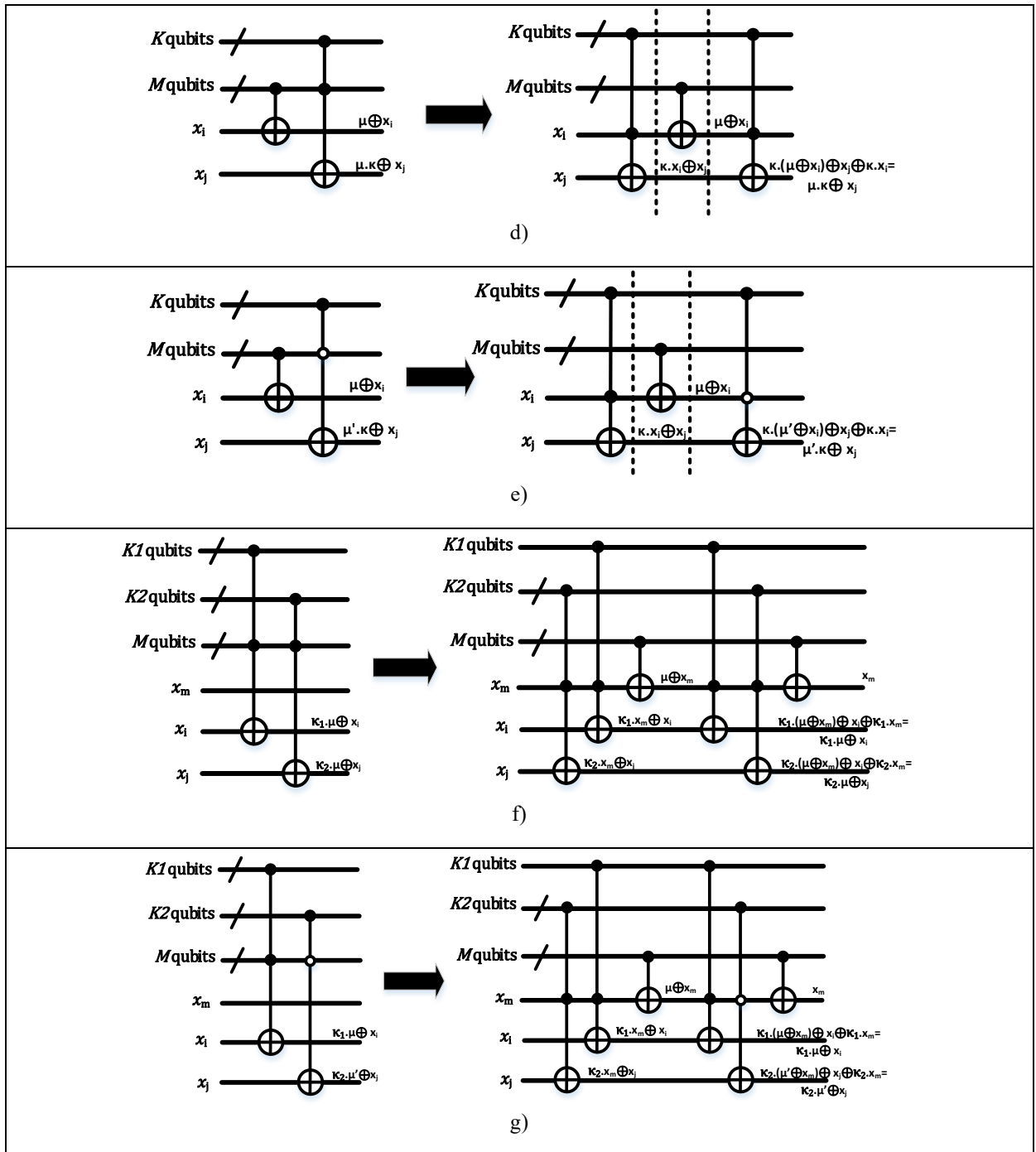

Figure 9. Factorizing common product terms with the proof of the templates (continued).

## 5. Scalability of the proposed method for functions with more variables

For functions with a larger number of variables, we propose a method which decomposes each vector $V_i$ into subfunctions which serve as building blocks of the main function. Each subfunction is derived from a subset of the true minterms corresponding to the elements of

$V_i$, exhibiting repetitive patterns. Once these subfunctions are identified, the proposed MQM method—comprising swapping and evaluation processes—can be independently applied to each subfunction. This technique has the potential to substantially reduce the computational complexity of the synthesis algorithm for these functions.

For a function of $n$ variables $x_1, \ldots, x_{Mid}, \ldots, x_n$, the vector $V_i$ contains the indices of the true minterms of the function $v_i$ with the main variable $i$. In this vector, which is sorted ascendingly, we look for consecutive indices with repetitive patterns of $x_{Mid} \ldots x_n$ of the indices and identical values of $x_1 \ldots x_{Mid-1}$ for each pattern. Consider a subset of consecutive indices with $p$ members and $k$ different patterns ($k = 1,2,\ldots$), $H_{j,k} = \{(\alpha_{Mid,1} \ldots \alpha_{n,1}), (\alpha_{Mid,2} \ldots \alpha_{n,2}), \ldots, (\alpha_{Mid,p} \ldots \alpha_{n,p})\}$, each appearing $q$ times in $V_i$ ($j = 1,\ldots,q$)). Then, the set of $H_j$'s with an identical sequence of $G = x_1 \ldots x_{Mid-1}$ is grouped together and is referred to as a $subfunction\_H_k$. Based on the number of iterations for each subfunction, denoted as $q$, each subfunction includes a set of indices $\{G_1, \ldots, G_q\}$ that collectively form a group referred to as $Group_k$. In fact, in each subfunction, $H$ represents the variable part of the indices while $G$ corresponds to the fixed or shared part. In each iteration of a subfunction the $H$ component remains constant, whereas the $G$ component changes. The members of $Group_k$ correspond to the true minterms of a function. For simplicity, we use this set and the corresponding function interchangeably. The same applies to $subfunction\_H_k$ and its function.

The final expression obtained for the vector $V_i$ will be as the following:

$$\bigoplus_k (Group_k . subfunction\_H_k)$$

where $k$ refers to the indices of the various patterns/subfunctions that exist in the vector $V_i$.

We can now apply the MQM and evaluation/swapping steps with the two subsets of variables separately (instead of $n$ variables) to find the circuits of $f_i$ more efficiently. The following pseudocode generally illustrates the process of generating groups and their corresponding subfunctions.

**Remark**: Let's consider vector $V_i$ related to a main variable $x_i$. Assume this vector contains $subfunction\_H_1$ and $subfunction\_H_2$ and also $Group$ 1 and $Group$ 2. If

$$subfunction\ H_1 = not(subfunction\ H_2)$$

and

$$Group\ 1 = not(Group\ 2)$$

then we will have:

$$subfunction\ H_1.Group\ 1 + subfunction\ H_1'.Group\ 1'$$
$$= (subfunction\ H_2 \oplus Group 1)$$
$$= (subfunction\ H_1 \oplus Group\ 2)$$

Therefore, the proposed algorithm only needs to be applied to the two functons $subfunction\ H_2$ and $Group\ 1$ or $subfunction\ H_1$ and $Group\ 2$.

Figures 10 and 11 show an example of a 7-variable function of the benchmark alu-bdd_288 from Revlib [27]. In this function, only vectors $V_6$ and $V_7$ are not empty. Figure 10 shows the indices of the vector $V_6$. The table highlights two patterns of 3-bit sequences, marked by red and green boxes (referred to as $subfunction\_H_1$ and $subfunction\_H_2$), which appear repeatedly within the vector. Each $subfunction$ contains a $Group$ consisting of 8 members, as detailed below:

| $subfunction\ H_1 = \{001,011,101,111\}$ |
| --- |
| $Group_1 = \{0000,0001,0011,0100,1010,1101,,1110,1111\}$ |
| $subfunction\ H_2 = \{000,010,100,110\}$ |
| $Group_2 = \{0010,0101,0110,0111,1000,1001,1011,1100\}$ |

Next, the conditions presented in the remark are true for $subfunction\_H$s and $Group$s. In this case, we need only apply the MQM algorithm and evaluation/swapping steps to $subfunction\_H_2$ and $Group\ 1$. The final expression is as follows:

$$(subfunction\_H_2 \oplus Group\ 1) \oplus x_i$$

where $x_i$ is the main variable. After applying the proposed algorithm, the above expression can be implemented as follows:

$$Toff(x_2', x_4, x_7), Toff(x_3', x_4', x_7), CNOT(x_1', x_7), CNOT(x_7, x_6)$$

To complete the synthesis process of the ALU benchmark, the values of the vector $V_7$ are also shown in Figure 11. The vector again consists of two patterns of 3-bit sequences, $subfunction\_H_1$ and $subfunction\_H_2$, indicated in red and green colors, respectively.

Therefore,

| $subfunction\_H_1 = \{000,001,010,011,100,101\}$ |
| --- |
| $Group_1 = \{0000,0001,0010,0011,0100,0110,,1001,1011\}$ |
| $subfunction\_H_2 = \{110,111\}$ |
| $Group_2 = \{0101,0111,1000,1010,1100,1101,1110,1111\}$ |

According to the remark, the conditions are true for $subfunction\ H_2 \oplus Group\ 1$. Therefore, in this case, we will have:

$$(subfunction\_H_2 \oplus Group\ 1) \oplus x_i$$

After applying the proposed algorithm, the above expression can be implemented as follows:

$$X(x_7), Toff(x_5, x_6, x_7), CNOT(x_1', x_2), CNOT(x_1, x_7), Toff(x_2', x_4, x_7), CNOT(x_1', x_2)$$

| $x_1x_2x_3x_4x_5x_6x_7$ | $x_1x_2x_3x_4x_5x_6x_7$ | $x_1x_2x_3x_4x_5x_6x_7$ | $x_1x_2x_3x_4x_5x_6x_7$ |
|---|---|---|---|
| 0000001 | 0101000 | 1010001 | 1111001 |
| 0000011 | 0101010 | 1010011 | 1111011 |
| 0000101 | 0101100 | 1010101 | 1111101 |
| 0000111 | 0101110 | 1010111 | 1111111 |
| 0001001 | 0110000 | 1011000 | |
| 0001011 | 0110010 | 1011010 | |
| 0001101 | 0110100 | 1011100 | |
| 0001111 | 0110110 | 1011110 | |
| 0010000 | 0111000 | 1100000 | |
| 0010010 | 0111010 | 1100010 | |
| 0010100 | 0111100 | 1100100 | |
| 0010110 | 0111110 | 1100110 | |
| 0011001 | 1000000 | 1101001 | |
| 0011011 | 1000010 | 1101011 | |
| 0011101 | 1000100 | 1101101 | |
| 0011111 | 1000110 | 1101111 | |
| 0100001 | 1001000 | 1110001 | |
| 0100011 | 1001010 | 1110011 | |
| 0100101 | 1001100 | 1110101 | |
| 0100111 | 1001110 | 1110111 | |

Figure 10: Illustration of vector $V_6$ related to benchmark alu-bdd_288.

In conclusion, by arranging the two expressions above in sequence, the final result gives us the synthesized circuit that is equivalent to the initial transformation matrix.

$$Toff(x_2', x_4, x_7), Toff(x_3', x_4', x_7), CNOT(x_1', x_7), CNOT(x_7, x_6), X(x_7),$$

$$Toff(x_5, x_6, x_7), CNOT(x_1', x_2), CNOT(1,7), Toff(2', 4, 7), CNOT(1', 2)$$

## 6. Experimental results

In this section, we compare our synthesis method with two references in Table 1. The first one is the Revlib site [27], known as a source for benchmarks in the reversible circuit synthesis field. The second one is one of the most recent papers on the exact synthesis of quantum circuits [4]. We use the benchmarks presented in [27], which are reversible circuits from the NCT library, to compare the results of our approach to the best reported ones in Revlib and those reported in [4].

| $x_1x_2x_3x_4x_5x_6x_7$ | $x_1x_2x_3x_4x_5x_6x_7$ | $x_1x_2x_3x_4x_5x_6x_7$ |
|:---:|:---:|:---:|
| 0000000 | 0100000 | 1010110 |
| 0000001 | 0100001 | 1010111 |
| 0000010 | 0100010 | 1011000 |
| 0000011 | 0100011 | 1011001 |
| 0000100 | 0100100 | 1011010 |
| 0000101 | 0100101 | 1011011 |
| 0001000 | 0101110 | 1011100 |
| 0001001 | 0101111 | 1011101 |
| 0001010 | 0110000 | 1100110 |
| 0001011 | 0110001 | 1100111 |
| 0001100 | 0110010 | 1101110 |
| 0001101 | 0110011 | 1101111 |
| 0010000 | 0110100 | 1110110 |
| 0010001 | 0110101 | 1110111 |
| 0010010 | 0111110 | 1111110 |
| 0010011 | 0111111 | 1111111 |
| 0010100 | 1000110 | |
| 0010101 | 1000111 | |
| 0011000 | 1001000 | |
| 0011001 | 1001001 | |
| 0011010 | 1001010 | |
| 0011011 | 1001011 | |
| 0011100 | 1001100 | |
| 0011101 | 1001101 | |

Figure 11: The illustration of a vector $V_7$ for ALU benchmark

However, the comparison made in [4] is based only on the number of T-gate levels, which is not a sufficient cost metric to compare the approaches. Therefore, in Table 1, we present a comparison in terms of gate count as well. To compare with [4], we need to synthesize multi-qubit control gates. To this end, the authors of [4] used references [5,28] and presented the synthesis results of multi-qubit control gates in terms of the number of T levels. We use those results as a reference for comparing our method with them [4]. Moreover, the circuits synthesized by our algorithm are presented in the last column of Table 1.

Our results confirm significant reductions in the costs which are the consequence of the global view by the proposed approach in the optimization process instead of the local view on CNOT gates. Using the MQM method, we can collect control qubits of gates and then factorize them to simplify the applied gates. Moreover, the templates presented in Figure 9 can consider the relations between outputs to simplify the circuit accordingly. This approach reduces the number of control qubits required for CNOT gates, whereby decreasing both the number of T gates and the depth of T-gate levels. It is suitable for current NISQ hardwares as well as error correction architectures such as surface codes which can apply the CNOT gates with the same control qubits or the same target qubits [29,30].

Besides, the results show that our decomposition technique enables the approach to synthesize functions with large number of inputs and achieve significant results. Unlike previous works that use data structures such as QMDD to make their methods more scalable, our approach directly exploits the patterns of true minterms of the functions. It means the algorithm focuses only on input-output qubits that differ and attempts to equalize them.

## 7. Conclusion

In this paper, we presented a transformation-based method for synthesizing reversible circuits with a high number of inputs. This approach employs a global view rather than the traditional local view. By using a modified version of the Quine-McCluskey method, we aggregate the control qubits of the involved gates into a single qubit. The circuit is then simplified through factoring and re-evaluating these control qubits.

The scalability of the proposed method allows it to handle circuits with a large number of inputs. Importantly, the algorithm does not require ancilla qubits as it uses the qubits of the functions to be synthesized. This feature, combined with the reduction in the number of control qubits and, consequently T gates, facilitates the effective implementation of circuits on quantum devices using error-correcting surface codes.

Since our method is exact, the results were compared with recent exact synthesis approaches [4]. They were also compared with the best results of the meta-heuristic work in [27]. The results demonstrate significant practical improvements.

Our future research will explore several directions. First, we will investigate methods for synthesizing multi-qubit CNOT gates and converting them into Toffoli gates. Additionally, our scalable proposed method can be employed for distributed quantum computing.

Furthermore, this work is part of a larger project in which we aim to synthesize quantum circuits with a quantum error correction module, particularly utilizing surface codes. We will continue our synthesis flow with this component in future research.

Table 1: Comparison of the proposed method with [23, 25].

| benchmark | #qubits | Revlib [27] | | Paper [4] | The proposed method | | |
|---|---|---|---|---|---|---|---|
| | | # applied operations | T-levels cost | T-levels cost | # applied operations | T-levels cost | Our results |
| 4mod5-bdd_287 | 7 | CNOT #3 Toffoli #4 | 8 | 160 | CNOT #4 Toffoli #3 | 6 | CNOT(4,5), CNOT(2,5), CNOT(5',7), CNOT(7,6), Toff(1,5,7), Toff(3',7,6), Toff(3,6,7) |
| alu-bdd_288 | 7 | CNOT #3 Toffoli #5 | 10 | 412 | CNOT #4 Toffoli #4 | 8 | Toff(2',4,7), Toff(3',4',7), CNOT(1',7), CNOT(7,6), X(7), Toff(5,6,7), CNOT(1',2), CNOT(1,7), Toff(2',4,7), CNOT(1',2) |
| f2_232 | 8 | Toffoli #3 3-qubit Toffoli #2 4-qubit Toffoli #8 | 286 | 264 | CNOT #10 Toffoli # 21 | 42 | CNOT(8,7),Toff(8,3,2),Toff(5,3,1),Toff(6',7,3),Toff(3,5,1),Toff(8,3,2), Toff(6',7,3),Toff(6,2,4),Toff(2,5,1),Toff(7',8',2),Toff(2,5,1),Toff(2,6,4),Toff(7',8',2), CNOT(6',7),Toff(3,6,4),Toff(3,8,2),Toff(5',7,3),Toff(3,8,2),Toff(3,6,4),Toff(5',7,3),CNOT(6',7), CNOT(7',8),CNOT(5,7),CNOT(5,6),Toff(6,7',5),,Toff(5',8',3),Toff(6,7',5), CNOT(5',6),CNOT(7,8),CNOT(8',7),CNOT(5,8),X(7),X(8) |
| rd53_251 | 8 | CNOT #4 Toffoli #4 3-qubit Toffoli #10 5-qubit Toffoli #2 | 264 | 1056 | CNOT #19 Toffoli #16 3-qubit Toffoli #1 | 44 | CNOT(7,8),CNOT(6',7),Toff(8,7,6),CNOT(4,5),CNOT(7',8),CNOT(6,8),CNOT(4,8),CNOT(8,1),CNOT(8,6), Toff(5,6,1),CNOT(8,6),CNOT(4,8),CNOT(8',6),Toff(6,7',8),Toff(5,8,1),Toff(6,7',8),Toff(8,7,6),X(6),CNOT(5,2), CNOT(6,2),CNOT(6',8),CNOT(8,2),Toff(5,6,2,3),Toff(7,8,2),CNOT(6',8),Toff(4,8,3),Toff(5,7,3), Toff(4,8,7),Toff(2,6,7),Toff(5,7,3),Toff(4,8,7),Toff(2,6,7),CNOT(4',5),Toff(6,7,8), CNOT(8',6),CNOT(6',7),Toff(6',7',8) |
| dc1_220 | 11 | Toffoli #11 3-qubit Toffoli #9 4-qubit Toffoli #9 | 418 | 412 | CNOT #19 Toffoli #30 | 60 | Toff(6,11',1),Toff(7,10',6),Toff(8,9,7),Toff(7',10',6),CNOT(10,9),CNOT(1,2),Toff(6',11',1),CNOT(1,2), CNOT(11',2),Toff(7,9',6),Toff(8,10,7),Toff(7',9',6),CNOT(9',10),Toff(2,10',3),Toff(6,11',2),Toff(2,10',3), CNOT(11',2),Toff(8,9',6),CNOT(5,4),CNOT(3,4),Toff(5,6',3),Toff(10',11',5),,Toff(5,6',3),CNOT(3,4), CNOT(5,4),Toff(10',11',5),Toff(8,9',6),CNOT(10,8),Toff(5,10',4),Toff(6',11',5),Toff(5,10',4),CNOT(11',5), Toff(8',9',6),Toff(6',11',5),CNOT(10,8),Toff(5,9',7),Toff(5,10',6),Toff(8',11',5),Toff(5,10',6),Toff(5,9',7), Toff(8',11',5),CNOT(9,10),CNOT(10,9),X9,X10,X11,CNOT(8,9),Toff(9,10',6),CNOT(8,9),CNOT(9,10), Toff(8,10,7),CNOT(9,10) |
| z4_268 | 11 | Toffoli #7 3-qubit Toffoli #10 4-qubit Toffoli #6 5-qubit Toffoli #8 | 870 | 3360 | CNOT#27 Toffoli #9 | 18 | CNOT(11,1),CNOT(8,1),CNOT(5,1),CNOT(11,6),CNOT(9,6),CNOT(5,8),Toff(6,8,2),CNOT(6',11), CNOT(11',5),Toff(5,8',2),CNOT(11',5),CNOT(5,6),Toff(6',8,5),CNOT(5,9),Toff(9,11,3),CNOT(5,3), CNOT(5,9),CNOT(7,3), CNOT, (10,3), CNOT(5,9),Toff(9,11,5),CNOT(7,10),Toff(5,10,4),Toff(7,10',4), Toff(9,11,5),CNOT(5,9),Toff(6',8,5),CNOT(6',9),CNOT(11,9),CNOT(9,6),CNOT(5,8), CNOT(6,11),CNOT(9,11),CNOT(11,9),CNOT(5,11),X5,X6,CNOT(7,10), X7 |
| cm152a_212 | 12 | 3-qubit Toffoli #3 4-qubit Toffoli #8 | 292 | 256 | CNOT#11 Toffoli #20 | 40 | X(2),X(3),X(4),Toff(3',4',5),CNOT(12,8),Toff(2,8,12),Toff(12,5,1),Toff(3',4',5),Toff(12,5,1),Toff(2,8,12), CNOT(12,8),Toff(3,4',12),CNOT(10,6),Toff(2,6,10),Toff(12,10,1),CNOT(4,12),CNOT(3,12),CNOT(11,7), Toff(2,7,11),Toff(12,11,1),CNOT(4,12),CNOT(9,5),Toff(2,5,9),Toff(12,9,1),Toff(3,4,12),Toff(12,9,1), Toff(2,5,9),CNOT(9,5), Toff(12,11,1),Toff(2,7,11),CNOT(11,7),Toff(12,10,1),Toff(2,6,10), CNOT(10,6) |

**References**:

1. Nielsen, M. A. & Chuang, I. L. *Quantum Computation and Quantum Information* (Cambridge Univ. Press, 2010).
2. Shor, P. W. Algorithms for quantum computation: discrete logarithms and factoring. *Proc. 35th Annu. Symp. Found. Comput. Sci.* 124–134 (1994). https://doi.org/10.1109/SFCS.1994.365700
3. Grover, L. K. A fast quantum mechanical algorithm for database search. *Proc. 28th Annu. ACM Symp. Theory Comput. (STOC)* (1996). https://doi.org/10.1145/237814.237866
4. Niemann, P., Wille, R. & Drechsler, R. Advanced exact synthesis of Clifford+T circuits. *Quantum Inf. Process.* **19**, (2020).
5. Giles, B. & Selinger, P. Exact synthesis of multiqubit Clifford+T circuits. *Phys. Rev. A* **87**, 032332 (2013).
6. Golubitsky, O., Falconer, S. M. & Maslov, D. Synthesis of the optimal 4-bit reversible circuits. *Proc. 34th Design Autom. Conf.* **70**, 653–656 (2010). https://doi.org/10.1145/1837274.1837440
7. Große, D., Wille, R., Dueck, G. W. & Drechsler, R. Exact multiple-control Toffoli network synthesis with SAT techniques. *IEEE Trans. Comput.-Aided Design Integr. Circuits Syst.* **28**, 703–715 (2009). https://doi.org/10.1109/TCAD.2009.2017215
8. Wille, R., Soeken, M., Przigoda, N. & Drechsler, R. Exact synthesis of Toffoli gate circuits with negative control lines. *Proc. 42nd Int. Symp. Multiple-Valued Logic* 69–74 (2012). https://doi.org/10.1109/ISMVL.2012.71
9. Mishchenko, A. & Perkowski, M. Fast heuristic minimization of exclusive-sums-of-products. *PDXScholar* (2025).
10. Sasamal, T. N., Gaur, H. M., Singh, A. K. & Mohan, A. Reversible circuit synthesis using evolutionary algorithms. *Lect. Notes Elect. Eng.* 115–128 (2019). https://doi.org/10.1007/978-981-13-8821-7_7
11. Saeedi, M., Zamani, M. S., Sedighi, M. & Sasanian, Z. Reversible circuit synthesis using a cycle-based approach. *ACM J. Emerg. Technol. Comput. Syst.* **6**, 1–26 (2010). https://doi.org/10.1145/1877745.1877747
12. Sasanian, Z., Saeedi, M., Sedighi, M. & Zamani, M. S. A cycle-based synthesis algorithm for reversible logic. *Proc. Asia South Pacific Design Autom. Conf.* 745–750 (2009). https://doi.org/10.1109/ASPDAC.2009.4796569
13. Arabzadeh, M., Saheb Zamani, M., Sedighi, M. & Saeedi, M. Depth-optimized reversible circuit synthesis. *Quantum Inf. Process.* **12**, 1677–1699 (2012). https://doi.org/10.1007/s11128-012-0482-8
14. Saeedi, M., Arabzadeh, M., Saheb Zamani, M. & Sedighi, M. Block-based quantum-logic synthesis. *Quantum Inf. Comput.* **11**, 362–378 (2011).

15. Maslov, D., Dueck, G. W. & Miller, D. M. Techniques for the synthesis of reversible Toffoli networks. *ACM Trans. Design Autom. Electron. Syst.* **12**, 42–42 (2007). https://doi.org/10.1145/1278349.1278355
16. Drechsler, R., Finder, A. & Wille, R. Improving ESOP-based synthesis of reversible logic using evolutionary algorithms. *Proc. Eur. Conf. Appl. Evol. Comput.* (Springer, 2011).
17. Bandyopadhyay, C., Rahaman, H. & Drechsler, R. A cube pairing approach for synthesis of ESOP-based reversible circuit. *Proc. 44th Int. Symp. Multiple-Valued Logic* (2014). https://doi.org/10.1109/ISMVL.2014
18. Datta, K. *et al.* An improved reversible circuit synthesis approach using clustering of ESOP cubes. *ACM J. Emerg. Technol. Comput. Syst.* **10**, 19 (2014).
19. Bandyopadhyay, C., Parekh, S., Roy, D. & Rahaman, H. Improving the designs of ESOP-based reversible circuits. *Lect. Notes Elect. Eng.* (2020).
20. Datta, K. *et al.* An ESOP-based reversible circuit synthesis flow using simulated annealing. In *Appl. Comput. Security Syst.: Vol. Two* (2015).
21. Wille, R. & Drechsler, R. BDD-based synthesis of reversible logic for large functions. *Proc. 46th Annu. Design Autom. Conf.* (2009).
22. Stojković, S. *et al.* Reversible circuits synthesis from functional decision diagrams by using node dependency matrices. *J. Circuits, Syst. Comput.* **29**, 2050192 (2020).
23. Law, J. J. & Rice, J. E. Line reduction in reversible circuits using KFDDs. *Proc. IEEE Pacific Rim Conf. Commun., Comput. Signal Process. (PACRIM)* (2015).
24. Niemann, P. *et al.* QMDDs: Efficient quantum function representation and manipulation. *IEEE Trans. Comput.-Aided Design Integr. Circuits Syst.* **34**, 1021–1031 (2015).
25. Zulehner, A. & Wille, R. Improving synthesis of reversible circuits: Exploiting redundancies in paths and nodes of QMDDs. *Proc. 9th Int. Conf. Reversible Comput. (RC)* (2017).
26. Ketineni, H. & Perkowski, M. Quantum algorithm for exact minimal ESOP minimization of incompletely specified Boolean functions and reversible synthesis. *Quantum Inf. Comput.* **23**, 1–21 (2023).
27. Wille, R. *et al.* RevLib: An online resource for reversible functions and reversible circuits. *Proc. 38th Int. Symp. Multiple-Valued Logic (ISMVL)* (2008). [Online]. Available: http://www.revlib.org/function_details.php?id=10
28. Amy, M. *et al.* A meet-in-the-middle algorithm for fast synthesis of depth-optimal quantum circuits. *IEEE Trans. Comput.-Aided Design Integr. Circuits Syst.* **32**, 818–830 (2013).
29. Herr, D., Nori, F. & Devitt, S. J. Lattice surgery translation for quantum computation. *New J. Phys.* **19**, 013034 (2017).
30. Litinski, D. & von Oppen, F. Lattice surgery with a twist: Simplifying Clifford gates of surface codes. *Quantum* **2**, 62 (2018).